\documentclass[fleqn,a4paper,11pt]{article}
\pdfoutput=1
\usepackage[ascii]{inputenc}
\usepackage[T1]{fontenc}
\usepackage[USenglish]{babel}
\usepackage{fullpage}
\usepackage{amsmath}
\usepackage{amssymb}
\usepackage{makeidx}
    \makeindex
\usepackage{times}
\usepackage{booktabs}
\usepackage{color}
\usepackage{authblk}
\usepackage[all]{xy}

\usepackage[pdftex]{epsfig}
\usepackage[pdftex,bookmarksnumbered]{hyperref}
\DeclareGraphicsExtensions{.pdf}

\chardef\bslash=`\\ 

\newcommand{\bibtex}{\ifx\is@itshape\f@shape{\fontshape{scit}\selectfont
Bib}\else\textsc{Bib}\fi\kern-.1em\TeX}

\newcommand{\ket}[1]{\lvert#1\rangle}                  
\newcommand{\bra}[1]{\langle#1\rvert}                  

\newcommand{\id}{{\mathbb{I}}}

\newcommand{\cc}{{\mathbb{C}}}
\newcommand{\tr}{\operatorname{tr}}
\makeatletter
\newcommand*{\overbar}[1]{$\overline{\hbox{#1}}\m@th$}
\makeatother
\newcommand{\ot}{\otimes}

\title{Multi-partite entanglement}\index{entanglement!multi-partite}
\author[1]{Michael Walter\thanks{michael.walter@stanford.edu}}
\author[2,3]{David Gross\thanks{david.gross@thp.uni-koeln.de}}
\author[4]{Jens Eisert\thanks{jense@physik.fu-berlin.de}}
\affil[1]{Stanford Institute for Theoretical Physics, Stanford University, Stanford, CA 94305, USA}
\affil[2]{Institute for Theoretical Physics, University of Cologne, 50937 K\"oln, Germany}
\affil[3]{Centre for Engineered Quantum Systems, School of Physics, The University of Sydney, Sydney, NSW 2006, Australia}
\affil[4]{Dahlem Center for Complex Quantum Systems, Freie Universit\"at Berlin, 14195 Berlin, Germany}
\date{}
\hypersetup{pdftitle={Multi-partite entanglement},pdfauthor={Michael Walter, David Gross, Jens Eisert}}

\begin{document}
\maketitle
\tableofcontents

\section{Introduction}

In this chapter, we generalize entanglement theory from two-partite to multi-partite systems.
The term \emph{multi-partite} may refer here to quantum systems composed of a macroscopic number of subsystems, such as the parts of an interacting many-body system as studied in condensed matter physics, or it may mean merely ``three''.
In this more general setting, it is still true that \emph{entanglement} refers to nonlocal properties of quantum states that cannot be explained classically.

It is immediately plausible that the phenomenology is much richer in the multi-partite regime.
Take the word ``locality'' for example.
There are super-exponentially many ways
to partition the constituents of an $N$-partite quantum system into non-overlapping groups, with each such partitioning giving rise to a legitimate locality constraint.
As a result of this complexity, the theory of multi-partite entanglement is much less \emph{canonical} than the
bipartite version.
By this, we mean that in the two-party case, questions like ``What is a natural unit of entanglement?'' or ``When is a state
maximally entangled'' tend to have unambiguous natural answers.
This is always never true for more than two subsystems, as we will see time and again in this chapter.

When the first version of this chapter was written a decade ago, it was common to say that a complete understanding of multi-partite entanglement had not \emph{yet} been reached.
Ten years later, many more multi-partite phenomena have been studied in detail, but we are arguably still far from a coherent picture.
We may thus need to come to terms with the fact that a canonical theory of multi-partite entanglement may not exist.
For example, as we will recall below, there is a basically unique way of quantifying bipartite pure state entanglement, whereas the ``right'' multi-partite measure strongly depends on the intended use of the entangled states.
The structure of this chapter reflects this multi-faceted aspect of multi-partite entanglement:

In Section~\ref{sec:general}, we begin by describing the ``coordinate system'' of the field:
Are we dealing with pure or mixed states; with single or multiple
copies; what notion of ``locality'' is being used;
do we aim to classify states according to their ``type of entanglement'' or to quantify it; etc.
In Section~\ref{sec:importantstates}, we describe important classes of multi-partite quantum states that admit an efficient classical description -- including matrix-product states, stabilizer states and bosonic and fermionic Gaussian states -- and we sketch their roles in quantum information theory.
Lastly, in Section~\ref{sec:specialized} we survey a variety of very different and largely independent aspects of multi-partite entanglement.
The purpose of these examples is to give a feeling for the breadth of the field.
Necessarily, we treat only a highly incomplete set of phenomena and the selection is necessarily subjective.
For a more exhaustive overview of the various aspects touched upon in this book chapter, we refer the readers to the excellent related review articles that have appeared since the first edition of this book~\cite{HorodeckiEntanglement,amico2008entanglement,guhne2009entanglement}.

One of the developments in the theory of multi-partite entanglement that has taken place since the first version of this chapter has been the significant deepening of connections between quantum information and condensed matter theory.
For example, it has become clear that notions of entanglement provide a fresh perspective to capture quantum many-body systems and phases of matter~\cite{AreaReview,amico2008entanglement,Orus-AnnPhys-2014,zeng2015quantum}.
We can mention these developments only briefly here -- but we expect that this connection will be a driving force behind the further development of multi-partite entanglement theory for the foreseeable future.

\section{General theory}
\label{sec:general}

\subsection{Classifying pure state entanglement}\label{mpePureStates}

We will start by considering multi-particle entanglement of \emph{pure} quantum states.
This is the study of state vectors in a Hilbert space
\begin{equation*}
  {\cal H}= {\cal H}_1\otimes {\cal H}_2\otimes\dots\otimes{\cal H}_N
\end{equation*}
of a quantum mechanical system of $N$ distinguishable constituents.
We assume that each particle is associated with a finite-dimensional Hilbert space ${\cal H}_i \simeq \mathbb{C}^{d_i}$, for some $d_i < \infty$.
The study of entanglement in systems of \emph{indistinguishable} particles has also gained increased attention (see,
e.g., Refs.~\cite{amico2008entanglement,
HorodeckiEntanglement,
PhysRevA.86.012337,
eckert_quantum_2002,
verstraete_quantum_2003,
schuch_quantum_2004,
PhysRevA.73.052323,
banuls_entanglement_2007,
franco_quantum_2016} and references therein)
but will not be covered in this introductory chapter.

The starting point of entanglement theory is to define the set of \emph{unentangled} states.
In the present setting, this corresponds to \emph{product states}\index{product state}, i.e.\ to vectors of the type
\begin{equation}\label{eqn:factor}
	\ket\psi=\ket{\psi^{(1)}} \otimes \dots \otimes \ket{\psi^{(N)}}.
\end{equation}
A state vector is \emph{entangled}\index{entangled pure state} if it is not of this form.
Entangled vectors are themselves grouped into different classes of ``equivalent'' entanglement.
There are various meaningful equivalence relations that give rise to different classifications. We will introduce two of them -- \emph{local unitary} equivalence and equivalence under \emph{local operations and classical communication} -- in the following.

\subsection{Local unitary equivalence}
\label{subsec:lu}

The finest distinction is the one based on \emph{local unitary (LU)}\index{LU} equivalence.
Here, two state vectors $\ket\psi,\ket\phi$ are considered equivalently entangled if they differ only by a local unitary basis change:
\begin{equation*}
	\ket\psi \sim_{LU} \ket\phi
	\quad
	\Leftrightarrow
	\quad
	\ket\psi = (U_1\otimes \dots \otimes U_N)\,\ket\phi
\end{equation*}
for suitable  $(d_i \times d_i)$ unitary matrices $U_i$.

To get a feeling for this classification, it is instructive to compare it to the bipartite case $N=2$.
For convenience, we assume that $d_1=d_2=d$ in what follows.
Then the \emph{Schmidt decomposition}\index{Schmidt!decomposition} says that for every vector $\ket\psi\in{\cal H}$, there are $d$ non-negative numbers $\{p_i\}$ summing to one (the \emph{Schmidt coefficients}\index{Schmidt!coefficients} or \emph{entanglement spectrum}\index{entanglement!spectrum}), as well as ortho-normal bases $\{\ket{\alpha_i}\}_{i=1}^d$ of ${\cal H}_1$ and $\{\ket{\beta_i}\}_{i=1}^d$ of ${\cal H}_2$ such that
\begin{equation}\label{eq:schmidt}
	\ket\psi = \sum_{i=1}^d \sqrt{p_i} \, \ket{\alpha_i} \otimes \ket{\beta_i}.
\end{equation}
Clearly, local unitary operations just change the bases, while keeping the $p_i$'s invariant
\begin{equation*}
	(U_1\otimes U_2)\,\ket\psi = \sum_{i=1}^d \sqrt{p_i} \, (U_1 \ket{\alpha_i}) \otimes (U_2\ket{\beta_i}).
\end{equation*}
Because any ortho-normal basis can be mapped onto any other such basis by a unitary operation, one concludes that two bipartite state vectors are LU-equivalent if and only if their Schmidt coefficients coincide.

This is a very satisfactory result for several reasons.
First, it gives a \emph{concise} answer to the entanglement classification problem.
Second, the Schmidt coefficients have a simple physical meaning: They are exactly the set of eigenvalues of each of the reduced density matrices
\begin{equation*}
	\rho^{(1)} = \tr_2 \ket\psi\bra\psi, \qquad
	\rho^{(2)} = \tr_1 \ket\psi\bra\psi.
\end{equation*}
As such, they can be estimated physically, e.g.\ using \emph{quantum state tomography} or direct \emph{spectrum estimation}\index{spectrum estimation} methods~\cite{alicki1988symmetry,keyl2001estimating,hayashi2002quantum,christandl2006spectra}.
Third, there is a simple and instructive proof of the validity of the Schmidt form involving just linear algebra.
We will not state it here in detail, but the idea is to expand $\ket\psi$ with respect to a product basis as
$\ket\psi = \sum_{i,j=1}^d T_{i,j} \ket i \otimes \ket j$.
Then the Schmidt coefficients are just the squared \emph{singular values} of the coefficient matrix $T_{i,j}$.
Unfortunately (and ominously), none of the three desirable properties we have identified in the bipartite case generalize to $N>2$ subsystems.

To show that there cannot be a concise label in the same
sense as above for LU-equivalence classes -- even for the supposedly simplest case of $N$ qubits ($d_i=2$) -- it suffices to count parameters:
Disregarding a global phase, it takes $2^{N+1}-2$ real parameters to specify a normalized quantum state in $\mathcal{H}=(\cc^2)^{\otimes N}$.
The group of local unitary transformations $\operatorname{SU}(2) \times \cdots \times \operatorname{SU}(2)$ on the other hand has $3N$ real parameters.
Because the set of state vectors that are LU equivalent to a given $\ket\psi$ is
the same as the image of $\ket\psi$ under all local unitaries, the dimension of an equivalence class cannot exceed $3N$ (it can be less if $\ket\psi$ is \emph{stabilized} by a continuous subset of the local unitaries).
Therefore, one needs at least $2^{N+1}-3N-2$ real numbers to parameterize the sets of inequivalent pure quantum states of $N$ qubits~\cite{LindenMultiEntanglement,CarteretMultiSchmidt}.
Second, it seems less clear what the immediate physical interpretation is for most of these exponentially many parameters.
Third, questions about bipartite entanglement can often be translated to matrix problems, as sketched above.
This makes the extremely rich toolbox of linear algebra applicable to bipartite entanglement theory.
In contrast, many-body pure states are mathematically described by \emph{tensors}\index{tensor}, with one index for each subsystem.
While tensors have also been studied extensively in pure mathematics~\cite{landsberg2012tensors}, the theory is much more challenging and less developed than linear algebra.
The aptly titled publication~\cite{lekheng} should serve as a well-informed warning.

Still, in particular for low-dimensional cases, many results have been obtained~\cite{LindenMultiEntanglement,
CarteretMultiSchmidt,GrasslInvariants,RainsInvariants,AcinThreeSchmidt1,briand2003complete,leifer2004measuring,luque2007unitary,PhysRevLett.104.020504}.
A systematic strategy is to look for \emph{invariants}\index{invariants} and \emph{normal forms}\index{normal form} under the group of local transformations.
From a mathematical point, these concepts are studied in the field of \emph{algebraic geometry}.
An \emph{invariant} is a function of the state vectors that does not change as we apply local unitary basis changes.
A simple example of invariants is given by the set of eigenvalues of the $N$ single-party reduced density matrices $\rho^{(i)} = \tr_{\setminus i} \ket\psi\bra\psi$ (this generalizes the bipartite Schmidt coefficients).
The number of independent invariants will be at least as large as the number of parameters identified above, i.e.\ exponential in the number of particles.
The idea behind \emph{normal forms}\index{normal form} is to look at the group of local basis changes as a ``gauge group'' that can be used to bring the state vectors into a distinguished form.
There must be only one such distinguished vector in every equivalence class -- so that two states are equivalent if and only if their respective normal forms coincide.
The two approaches are related: The parameters that appear in the normal form are, by definition, invariants.

To give a taste of these statements and the employed proof methods, we briefly present the derivation of a normal form for the simplest case -- three qubits~\cite{AcinThreeSchmidt1}.
We start with a general state vector
\begin{equation}\label{eqn:tensor_expansion}
  \ket\psi = \sum_{i,j,k} \alpha_{i,j,k} \ket{i,j,k}.
\end{equation}
Define two matrices $T_0$, $T_1$ by $(T_i)_{j,k} = \alpha_{i,j,k}$.
If we apply a unitary operator $U_1$ with matrix elements $u_{i,j}$
to the first qubit, then the matrix $T_0$ transforms according to
\begin{equation*}
  T_0' = u_{0,0} T_0 + u_{0,1} T_1.
\end{equation*}
From this transformation law, one can easily see that one may always choose $U_1$ such that $\det(T_0')=0$
(essentially because this conditions amounts to a quadratic equation in $u_{0,1}/ u_{0,0}$, which always has a solution over the complex numbers).
Next, if we apply unitaries $U_2, U_3$ to the second and third system respectively, the matrix $T_0'$ transforms according to
\begin{equation}\label{eqn:trans}
	T_0' \mapsto T_0'' = U_2 T_0 ' U_3^T.
\end{equation}
We can use this freedom to diagonalize $T_0''$ -- this is the singular value decomposition.
Since the determinant is still zero, there can be at most one non-zero singular value $\lambda_0$, so we arrive at the form
\begin{equation*}
  T''_0 =
    \left(\begin{array}{cc}
      \lambda_0& 0 \\
      0 & 0
    \end{array}\right), \qquad \lambda_0 \geq 0.
\end{equation*}
By the definition of $T_0$, this means that the coefficients of the transformed state fulfill $\alpha_{0,i,j} = \lambda_0 \delta_{i,0} \delta_{j,0}$, while the four coefficients $\alpha_{1,i,j}$ are arbitrary.
This form remains unchanged if we act on the state by diagonal unitaries $V_1, V_2, V_3$.
Choosing the $V_i$'s suitably allows us to make three of the four coefficients $\alpha_{1,i,j}$ real.
We are left with
\begin{equation}\label{ThreeSchmidtDecomposition}
\begin{aligned}
  &\quad (V_1 U_1 \otimes V_2 U_2 \otimes V_3 U_3)
  \ket\psi \\
  &= \lambda_0 \ket{0,0,0} + \lambda_1 e^{i\phi} \ket{1,0,0}
  + \lambda_2 \ket{1,0,1} + \lambda_3\ket{1,1,0} +
  \lambda_4\ket{1,1,1},
\end{aligned}
\end{equation}
with real coefficients $\lambda_i$. Normalization requires that $\sum_i
\lambda_i^2=1$. It is shown in Ref.~\cite{AcinThreeSchmidt1} that
$0\leq\phi\leq\pi$ can always be achieved and, further, that for a
generic\footnote{In this context \emph{generic} means for all state vectors
but a set of measure zero.} state vector the normal form in Eq.~\eqref{ThreeSchmidtDecomposition} is unique.
In accordance with the dimension formula that we derived earlier, it depends on five independent parameters.

The technique presented here has been extended to provide a normal form for pure states of $N$ qubit-systems for arbitrary $N$~\cite{PhysRevLett.104.020504}.
As in the $N=3$ example presented above, two generic vectors are LU-equivalent if and only if their normal forms coincide.
In accordance with our estimates above, the normal form of Ref.~\cite{PhysRevLett.104.020504} depends on exponentially many parameters.
While -- as expected -- the normal form does not identify a concise set of parameters labeling LU-equivalence classes, the mathematical framework can be very useful for the analysis of multi-partite entanglement.
For example, the follow-up paper Ref.~\cite{PhysRevA.82.032121} builds on these normal forms to construct pairs of multi-partite entangled states
with the property that all entropies of subsystems coincide, however, the states are not LU equivalent.

\subsection{Equivalence under local operations and classical communication}

\label{subsec:LOCCSLOCC}

From an operational perspective, LU equivalence is justified because we cannot create entangled states from separable states by local unitary basis changes alone.
This suggests that we can obtain coarser notions of ``equivalent entanglement'' by considering larger classes of operations that have this property.

The most natural such class consists of \emph{local operations and classical communication (LOCC)}\index{LOCC}.
LOCC protocols are best described in the ``distant laboratories model''~\cite{chitambar2014everything}.
Here, we imagine that each of the $N$ particles is held in its own laboratory.
The particles may have interacted in the past, so their joint state vector $\ket{\psi}\in{\cal H}$ may be entangled.
We assume that each laboratory is equipped to perform arbitrary experiments on the particle it controls and that the experimenters can coordinate their actions by exchanging classical information.
However, no quantum systems can be exchanged between laboratories.
It is not hard to verify that no entangled state can be created from a product state by LOCC alone.

An LOCC protocol proceeds in several \emph{rounds}.
In each round, one of the experimenters performs a (POVM) measurement on their particle.
They keep the post-measurement state and broadcast the classical outcome to the other laboratories.
In the next round, another party gets to act on their system with an operation that may depend on the previous measurement outcomes, and so on.
We say that two states are \emph{LOCC-equivalent} if they can be converted into each other by a protocol of this form.
There are several useful variants of this definition.
For example, two states are LOCC$_r$-equivalent if they can be converted into each other using an LOCC protocol with no more than $r$ rounds.
They are \overbar{LOCC}-equivalent if, starting from any of them, we can approximate the other one to arbitrary precision, as the number of rounds $r$ tends to infinity.
It also makes sense to define versions of these equivalence relations where the transformations need only succeed with some fixed probability of success.

While natural and physically well-defined, no tractable mathematical description of LOCC-equivalence in the multi-partite case has been identified so far.
For example, there is no known algorithm that decides whether two vectors are \overbar{LOCC} equivalent, even if one allows for, say, exponential runtime in the total dimension.
It seems conceivable that no such algorithm exists.
(In contrast, \emph{Nielsen's theorem}\index{Nielsen's theorem}~\cite{nielsen1999conditions} provides a simple criterion for the equivalence of bipartite pure states under LOCC.)
For an introduction into the structure of the LOCC-problem, see Ref.~\cite{chitambar2014everything}.
What can still be described, despite the difficulties of identifying LOCC-equivalent pure states in the multi-partite setting, is a set of states that are in a sense ``maximally useful'' under LOCC manipulation. This set, called the \emph{maximally entangled set}\index{maximally entangled set}~\cite{KrausMaximallyEntangled,Spee}, has the property that any state outside can be obtained via LOCC from one of the states in the set, and that no state included in the set can be obtained from any other state in the set via LOCC, in some analogy to the properties of a maximally entangled state in the bipartite setting.

There is a variant of LOCC with a less satisfactory physical interpretation, but a rather more tractable mathematical description.
It is the notion of \emph{stochastic LOCC (SLOCC)}\index{SLOCC}\index{LOCC!stochastic} equivalence.
Here we deem two states equivalent if they can be converted into each other by LOCC with some non-zero probability (which, however, may be tiny!).
As before, an SLOCC protocol consists of several rounds in each of which the parties perform operations on their respective systems, possibly depending on previous measurement results. One can think of the protocol as splitting into different \emph{branches} with each measurement.
A transformation $\ket\psi \rightarrow \ket\phi$ is possible if and only if at least \emph{one} of these branches does the job.
But the effect of each single branch can be described by one \emph{Kraus operator}\index{Kraus operator} $A_i$ for each system
as
\[ \ket\psi \mapsto \bigl( A_1\otimes\cdots\otimes A_N \bigr) \ket\psi. \]
Thus $\ket\psi \rightarrow \ket\phi$ is possible under SLOCC if there exists operators $A_i$ and a scalar $\lambda \in\cc$ such that
\begin{equation}\label{eqn:slocc_equiv}
	\bigl(A_1\otimes\cdots\otimes A_N\bigr) \ket\psi = \lambda \ket\phi.
\end{equation}
It is not difficult to prove that two states $\ket\psi, \ket\phi$ are SLOCC-equivalent if and only if Eq.~\eqref{eqn:slocc_equiv} can be realized with matrices that have unit determinant $\det A_i = 1$~\cite{DuerThreeQubits,Frank}.
Matrices with that property form a group -- the \emph{special linear group} $\operatorname{SL}$.
Thus, mathematically, two vectors are SLOCC-equivalent if and only if, up to normalization, they lie on a single orbit of the group $\operatorname{SL}(\cc^{d_1})\times\cdots\times\operatorname{SL}(\cc^{d_N})$.
We note that the term \emph{filtering operation}\index{filtering operation} is used synonymously with SLOCC.

Having thus established a framework for dealing with SLOCC operations, we can proceed as in the case of local unitary equivalence.
By simply substituting the local unitary group by
the group
$\operatorname{SL}(\cc^2)\times\cdots\times\operatorname{SL}(\cc^2)$,
one finds a lower bound of $2^{N+1}-6N-2$ parameters that are
necessary to label SLOCC equivalence classes of an $N$-qubit system.
Again, there has been considerable work on invariants and normal forms for small systems (see,
e.g., Refs.~\cite{DuerThreeQubits,Frank,briand2003complete,osterloh2005constructing,luque2006algebraic,osterloh2006entanglement,klyachko2007dynamical}).

For three qubits, the above formula gives no nontrivial lower bound on the number of parameters and therefore one might expect that there is only a discrete set of inequivalent classes.
This turns out to be true~\cite{DuerThreeQubits}:

First, we note that \emph{product states}\index{product state} $\ket{\psi^{(1)}}\ot\ket{\psi^{(2)}}\ot\ket{\psi^{(3)}}$ certainly form a class of their own, because local operations can never
create entanglement between previously unentangled systems.
For the same reason vectors of the form $\ket{\psi^{(1)}}\ot\ket{\Phi^{(2,3)}}$ with some non-factoring state vector $\ket{\Phi^{(2,3)}}$ constitute an SLOCC equivalence class, the class of \emph{bipartite} entangled states that factor with respect to the bipartition 1$\mid$23.
There are two other such bipartitions -- 2$\mid$13, and 3$\mid$12 -- and they give rise to in total three bipartite classes.
Calling these sets \emph{equivalence classes} is justified, because any two entangled pure states of two qubits are equivalent under SLOCC.
Finally, we are left with the set of \emph{genuinely entangled}\index{entanglement!genuine} vectors that admit no representation as tensor products.
Do they form a single equivalence class? It turns out that this is not the case.
This can be shown by identifying an SLOCC-invariant that takes different values on two suitable genuinely entangled states.
We will briefly describe two invariants -- the \emph{hyperdeterminant} and the \emph{tensor rank} -- that each do the job.

Before we define the hyperdeterminant, we first consider the bipartite case.
Let
\begin{equation}\label{eq:coeff matrix}
	\ket\psi = \sum_{i,j=1}^d T_{i,j} \ket i \otimes \ket j,
	\qquad
	\ket\phi = \sum_{i,j=1}^d T_{i,j}' \ket i \otimes \ket j
\end{equation}
be the expansion of two state vectors with respect to a product basis.
As noted before, the coefficients $T_{i,j}$ define a matrix $T$ -- and likewise for the primed version.
Now assume $\ket\psi$ and $\ket\phi$ are SLOCC-equivalent, i.e.\ there exist unit-determinant Kraus operators $A_1, A_2$ and a non-zero scalar $\lambda$ such that $A_1\otimes A_2 \ket\psi = \lambda \ket\phi$.
As in Eq.~\eqref{eqn:trans}, one then finds that
\begin{equation*}
	 \lambda T' = A_1 T A_2^T
\end{equation*}
and thus
\begin{equation*}
	\det \lambda T'
	=  \det A_1 T A_2^T = (\det A_1) (\det T) (\det A_2) = \det T.
\end{equation*}
There are two conclusions one can draw from this calculation.
First, the determinant of the coefficient matrix is invariant under unit-determinant local filtering operations.
Second, because $\det \lambda T'=\lambda^d \det T'$, it holds that either both $\det T$ and $\det T'$ are equal to zero, or neither is.
Thus, the property ``Is the determinant of coefficient matrix zero?'' is an SLOCC-invariant.
In particular, a pure state $\ket\psi$ of two qubits is entangled if and only if $\det T\neq0$, and otherwise a product state.

Caley's \emph{hyperdeterminant}\index{Caley's hyperdeterminant}\index{hyperdeterminant} is a generalization of the determinant to tensors in $\cc^2\otimes\cc^2\otimes\cc^2$.
There are various ways of expressing it, none of which are particularly transparent.
One definition goes as follows:
If $\ket\psi$ has expansion coefficients $\alpha_{i,j,k}$ as in Eq.~\eqref{eqn:tensor_expansion}, then
\begin{equation}\label{eq:hyperdet}
	\mathrm{Det}(\psi)
	=
	\alpha_{i_1,j_1,k_1}
	\alpha_{i_2,j_2,k_2}
	\alpha_{i_3,j_3,k_3}
	\alpha_{i_4,j_4,k_4}
	\epsilon_{i_1,i_2}
	\epsilon_{i_3,i_4}
	\epsilon_{j_1,j_2}
	\epsilon_{j_3,j_4}
	\epsilon_{k_1,k_2}
	\epsilon_{k_3,k_4},
\end{equation}
where $\epsilon_{i,j}$ is the completely antisymmetric (or Levi-Civita) tensor~\cite{verstraete_normal_2003}.
In any case, it is known that the hyperdeterminant is invariant under the local special linear group.
Arguing as above, this means that zero/non-zero values of the hyperdeterminant can be used to distinguish SLOCC-equivalent classes~\cite{miyake_classification_2003}.
One may now verify by direct calculation that $\mathrm{Det}(\mathrm{W}) = 0 \neq \mathrm{Det}(\mathrm{GHZ})$.
In this sense, there are two ``inequivalent forms'' of genuinely tripartite entanglement of three qubits~\cite{DuerThreeQubits}.

So far, we have only extracted binary labels (zero / non-zero) for SLOCC-classes from the determinant and the hyperdeterminant.
There is a quite general way to obtain numerical invariants.
Indeed, both are \emph{homogeneous} functions, which means that $f(\lambda \psi) = \lambda^d f(\psi)$ for any scalar $\lambda$ and some fixed integer $d$, known as the \emph{degree} of the function.
For example, it is plain from Eq.~\eqref{eq:hyperdet} that $\mathrm{Det}$ is a homogeneous function of degree 4 in the expansion coefficients $\alpha_{i,j,k}$.
More generally, if $f,f'$ are two $\operatorname{SL}$-invariants of degree $d,d'$ then their ratio $f^{d'}(\psi)/{f'}^d(\psi)$, is insensitive to a rescaling $\psi \mapsto \lambda \psi$ of the state.
The ratio is thus an invariant quantity on SLOCC classes.
Such numbers can be used to distinguish inequivalent types of multi-partite entanglement~\cite{GourWallach} -- but the numerical value itself does not carry an obvious interpretation.
In Section~\ref{QuantifyMulti}, we will see that homogeneous $\operatorname{SL}$-invariants can also be used to construct quantitative \emph{measures} of entanglement.

We now discuss another useful invariant that separates the two states, known as the tensor rank.
Any pure state vector can be written in the form
\begin{equation}\label{schmidtMeasure}
  \ket{\psi} = \sum_{i=1}^R c_i \ket{\psi_{i}^{(1)} }\ot\dots\ot\ket{\psi_{i}^{(N)}}.
\end{equation}
Unlike in the Schmidt decomposition~\eqref{eq:schmidt}, we do not
(and in fact cannot in general)
require the vectors $\{\ket{\psi_i^{(j)}}\}_{i=1}^{R}$ to be orthogonal for each subsystem.
Now let $R_{\min}(\psi)$ denote the minimal number of product terms needed to express $\ket\psi$.
This number is the \emph{tensor rank}\index{tensor rank}\index{rank!tensor} of $\ket\psi$. It naturally generalizes the matrix rank (relevant for $N=2$).
We will re-visit this invariant in Section~\ref{QuantifyMulti}, where its logarithm will be called the \emph{Schmidt measure}.
A moment's thought shows that the tensor rank is constant under invertible SLOCC operations.
Now consider the states vectors~\cite{greenberger1989going,DuerThreeQubits}
\begin{equation}\label{eq:GZH and W}
\begin{aligned}
  \ket{\mathrm{GHZ}} &= \frac1{\sqrt 2} \bigl( \ket{0,0,0} + \ket{1,1,1} \bigr),\\
  \ket{\mathrm{W}} &= \frac1{\sqrt 3} \bigl( \ket{0,0,1} + \ket{0,1,0 } + \ket{1,0,0} \bigr).
\end{aligned}
\end{equation}
As we will now show, there is no way of expressing $\ket{\mathrm{W}}$ using only two product terms.
The idea is to compare the range of the reduced density matrices $\rho^{(2,3)}$.
For the GHZ state,
\begin{equation}\label{eq:GHZ rdm}
	\rho^{(2,3)}=\tr_1 \ket{\mathrm{GHZ}}\bra{\mathrm{GHZ}} = \frac12 \bigl( \ket{0,0}\bra{0,0} + \ket{1,1}\bra{1,1} \bigr),
\end{equation}
and so the range contains at least two product vectors.
The same is true for all genuinely entangled states that can be written as a superposition of two product terms.
This follows from the easily established fact that all such states are SLOCC equivalent to the GHZ state.
However, the reduced density matrix of the W state is given by
\begin{equation}\label{eq:W rdm}
  \tr_1 \ket{\mathrm{W}}\bra{\mathrm{W}} =  \frac13 \ket{0,0}\bra{0,0} + \frac23 \ket{\Psi^+}\bra{\Psi^+},
\end{equation}
with $\ket{\Psi^+}=(\ket{0,1}+\ket{1,0})/\sqrt{2}$.
Its range consists of all vectors of the form $\alpha \ket{0,0} + \beta (\ket{0,1} + \ket{1,0})$.
As discussed above, such a vector is a product state if and only if the determinant $\det T$ of its coefficient matrix
\begin{equation}T = \left(\begin{array}{cc}\alpha & \beta\\\beta & 0\end{array}\right)\end{equation} is zero.
But $\det T = -\beta^2$, and hence there is only a single product state in the range of the reduced density matrix, namely $\ket{0,0}$ for $\beta=0$.
This shows that $\ket{\mathrm{W}}$ cannot be written using two product terms.
Since on the other hand we have the decompositions~\eqref{eq:GZH and W}, we conclude that $R_{\min}(\mathrm{W})=3$, while $R_{\min}(\mathrm{GHZ})=2$.
It follows that the two states cannot be converted into each other by SLOCC.
This provides an alternative proof of their inequivalence.%

\medskip

While GHZ and W states cannot be exactly transformed into each other with any probability of success,
it is true that the W state can be \emph{approximated} to arbitrary precision by states in the GHZ class.
The converse does not hold -- in this sense, GHZ states are \emph{more} entangled than the W states (even though the higher tensor rank of the W state might have suggested the contrary).
That W states can be approximated by GHZ-class states can be seen as follows~\cite{vrana2015asymptotic}:%
\[ \ket{\psi_\epsilon} := \frac 1 {\sqrt 3\,\varepsilon} \bigl( \ket 0 + \varepsilon \ket 1 \bigr)^{\ot 3} - \frac 1 {\sqrt 3\,\varepsilon} {\ket 0}^{\ot 3} = \ket{\mathrm{W}} + O(\varepsilon). \]
The state $\ket{\psi_\epsilon}$ consists of two product terms, and it can easily be obtained from $\ket{\mathrm{GHZ}}$ by an invertible SLOCC operation.
The fact that we can approximate the W state by states of smaller tensor rank is quite remarkable.
In contrast, matrix ranks can never increase when we take limits, suggesting that the geometry of tensors is rather more subtle~\cite{landsberg2012tensors}.

The above classification is complete: The three-qubit pure states are partitioned into a total of six SLOCC equivalence classes.
Three-qubit W-states and GHZ-states have been experimentally realized, both purely optically using postselection~\cite{Wein,Zeil} and in ion traps~\cite{Blatt}.

It is instructive to discuss some of the properties of the GHZ and W states.
After measuring the first qubit in the computational basis, $\ket{\mathrm{GHZ}}$ collapses into a product state vector on the systems labeled $2$ and $3$. In contrast, if we instead measure in the eigenbasis $(\ket 0\pm\ket 1)/\sqrt 2$ of the
Pauli $X$-observable, the GHZ state collapses into one of two maximally entangled Bell states on the remaining systems.
This leads to remarkable nonclassical correlations which we discuss further in Section~\ref{subsec:non-locality} below.
If we discard the measurement outcome, or if we simply trace out the first qubit of the GHZ state, the remaining systems will be described by the unentangled bipartite mixed state~\eqref{eq:GHZ rdm}.
In contrast, for $\ket{\mathrm{W}}$ the bipartite density matrix~\eqref{eq:W rdm} is entangled.
In this sense the entanglement of $\ket{\mathrm{W}}$ is more \emph{robust}\index{robustness of entanglement} under particle loss than the one of $\ket{\mathrm{GHZ}}$.

Can a ``maximally robust'' three-qubit state be conceived that leaves any pair of systems in a Bell state if the third particle is lost?
Unfortunately not, because if e.g.\ $\tr_1\ket\psi\bra\psi$ is maximally entangled then $\ket\psi$ necessarily factors between systems $1$ and $2,3$.
It follows that $\tr_2 \ket\psi\bra\psi$ and $\tr_3 \ket\psi\bra\psi$ are not just unentangled but in fact completely uncorrelated~\cite{Tangle}.
This important observation is known as the \emph{monogamous nature}\index{monogamy of entanglement} of entanglement.

The clear-cut characterization of three-qubit entanglement breaks down immediately if we consider more than three particles or higher dimensions.
Already for four qubits, there are infinitely many SLOCC equivalence classes~\cite{Frank}.
This inevitable explosion in complexity, as predicted by parameter counting, suggests a change in perspective.
Rather than aiming for an exhaustive classification, we may instead classify states according to their utility for particular tasks, or else base a classification on physically accessible data.
An example for the former point of view are the classes of \emph{tensor network states} for given \emph{bond dimensions}, as described in Section~\ref{sec:manybody}.
These states have an efficient description that is well-suited to simulate e.g.\ ground states of quantum many-body systems.
An example for the latter point of view are classifications of multi-partite pure states according to the information accessible from their single-body or few-body reduced density matrices~\cite{toth2007optimal,wurflinger2012nonlocal,laskowski2012incompatible,walter2013entanglement,sawicki2012critical,sawicki2014convexity,sawicki2013pure,walter2014phd}, which are easily accessible in experiments.

\subsection{Asymptotic manipulation of pure multi-partite quantum states}
\label{sec:asymptotic}

Instead of manipulating quantum systems at the level of single specimens, entanglement manipulation is also meaningful in the asymptotic limit.
Here, one assumes that one has many identically prepared systems at hand, in a state $\ket\psi^{\otimes n}$, and aims at transforming them into many other identical states $\ket\phi^{\otimes m}$, for large $n$ and $m$.
As before, the state vectors $\ket\psi,\ket\phi$ are elements of an $N$-partite quantum system, so that the total Hilbert space containing $\ket\psi^{\otimes n}$ is
\begin{eqnarray*}
  &&{\cal H}^{(1)}_1\otimes {\cal H}^{(1)}_2\otimes\dots\otimes{\cal H}^{(1)}_N \\
  &\otimes&{\cal H}^{(2)}_1\otimes {\cal H}^{(2)}_2\otimes\dots\otimes{\cal H}^{(2)}_N\\
  &\otimes&\qquad\vdots \\
  &\otimes&{\cal H}^{(n)}_1\otimes {\cal H}^{(n)}_2\otimes\dots\otimes{\cal H}^{(n)}_N.
\end{eqnarray*}
Thus, in the tensor product above, every row corresponds to one copy of $\ket\psi$ and every column to one of the $N$ parties.
As in Section~\ref{subsec:LOCCSLOCC}, we adopt the ``distant laboratory model'' and imagine that each column is held in one lab.
Our analysis will be based on LOCC transformations.
As before, \emph{local} operations are confined to one lab.
The difference is that each lab is now associated with a tensor product Hilbert space $\mathcal{H}^{(1)}_i \otimes \dots \otimes \mathcal{H}^{(m)}_i$ describing the copies and that collective operations on the copies are allowed.
It makes sense not to require that the target state is reached exactly, but only with an error that is asymptotically negligible.

It is instructive again to briefly reconsider the situation when
only two subsystems are present~\cite{Reversible,Uniqueness,BennettMulti,ReversibleVidal,brandao2016mathematics}.
There, it turns out that the basic unit of bipartite entanglement is the \emph{EPR pair}\index{EPR!pair}
\begin{equation}\label{eq:EPR}
	\ket{\mathrm{EPR}} = \frac1{\sqrt 2}\left( \ket{0,0} +\ket{1,1} \right).
\end{equation}
Indeed, if $\ket{\psi}$ is any bipartite pure state, then there exists a \emph{rate} $r$ such that the transformation
\begin{equation*}
	\ket\psi^{\otimes n} \to
	\ket{\mathrm{EPR}}^{\otimes \lfloor r \cdot n \rfloor}
\end{equation*}
is possible under LOCC with an approximation error that goes to $0$ as $n$ goes to infinity (here, $\lfloor x \rfloor$ is the largest integer smaller than $x$).
The transformation is \emph{reversible} in that
\begin{equation*}
	\ket{\mathrm{EPR}}^{\otimes \lfloor r \cdot n \rfloor}
	\to
	\ket\psi^{\otimes n}
\end{equation*}
is also realizable.
In this sense, there is only a single type of bipartite entanglement and $r$ quantifies ``how much'' of it is present in a given state.
What is more: there is even a simple formula for the rate $r$: It is given by the \emph{entropy of entanglement}\index{entropy!of entanglement} or \emph{entanglement entropy}\index{entanglement!entropy} $E(\psi)$, which is the Shannon entropy $H(p)$ of the Schmidt coefficients:
\begin{equation*}
	E(\psi) = H(p) = - \sum_{i} p_i \log p_i,
  \qquad
	\ket\psi = \sum_{i=1} \sqrt{p_i} \, \ket{\alpha_i} \otimes \ket{\beta_i}.
\end{equation*}
Equivalently, the entropy of entanglement of a pure state is equal the \emph{von Neumann entropy}\index{entropy!von Neumann} $S(\rho) = -\tr \rho \log \rho$ of either reduced density matrix:
\begin{equation*}
	E(\psi) = S(\rho^{(A)}) = S(\rho^{(B)}),
	\qquad
  \rho^{(A)}=\tr_B \ket\psi\bra\psi,
  \qquad
	\rho^{(B)}=\tr_A \ket\psi\bra\psi.
\end{equation*}

Again, it turns out that for more than two parties, the situation is much more complex
than before.
Before stating what the situation is like in the multi-partite setting, let us first make the concept of {\it asymptotic reversibility} more precise.
If $\rho^{\otimes n}$ can be transformed under LOCC into $\sigma^{\otimes m}$ to arbitrary fidelity, there is no reason why $n/m$ should be an integer.
So to simplify notation, one typically also takes non-integer yields into account.
One says that $\ket\psi^{\otimes x}$ is \emph{asymptotically reducible} to  $\ket{\phi}^{\otimes y}$ under LOCC, if for all $\delta,\varepsilon>0$ there exist natural $n,m$ such that
\begin{equation*}
  \left\lvert \frac{n}{m} - \frac{x}{y} \right\rvert < \delta,
  \quad
  \big\lVert \Psi( \ket\psi\bra\psi^{\otimes n}) - \ket\phi\bra\phi^{\otimes m} \big\rVert_1 > 1 - \varepsilon.
\end{equation*}
Here, $\lVert\cdot\rVert_1$ denotes the trace-norm as a distance measure, and $\Psi$ is quantum operation which is LOCC.
If both $\ket\psi^{\otimes x}$ can be transformed into
$\ket\phi^{\otimes y}$ as well as $\ket\phi^{\otimes y}$
into $\ket\psi^{\otimes x}$, the transformation is \emph{asymptotically reversible}.
Using this notation, in the bipartite case, it is always true that any $\ket\psi$ can asymptotically be transformed into
$\ket{\mathrm{EPR}}^{\otimes E(\psi) }$.

In the multi-partite setting, there is no analogue of the EPR state: There is no single state to which any other state can be asymptotically reversibly transformed.
A generalized notion is that of a \emph{minimal reversible entanglement generating set (MREGS)}\index{MREG}\index{minimal reversible entanglement generating set}.
An MREGS $S$ is a set of pure states such that any other state can be generated from $S$ by means of reversible asymptotic LOCC. It must be minimal in the sense that no set of smaller cardinality possesses the same property~\cite{BennettMulti,LindenMREGS,VirmaniMREGS}.

After this preparation, what is now the MREGS for, say, a
tri-partite quantum system?
Even in this relatively
simple case, no conclusive answer is known.
Only a few states have been identified that must be contained in any
MREGS.
At first one might be tempted
to think that three different maximally entangled qubit pairs,
shared by two systems each, already form an MREGS. This
natural conjecture is not immediately ruled out by what we have
seen in the previous subsection: after all, we do not aim at
transforming quantum states of single specimens, but rather allow
for asymptotic state manipulation. Yet, it can be shown that
merely to consider maximally entangled qubit pairs is not
sufficient to construct an MREGS~\cite{LindenMREGS}.
What is more, even
\begin{equation}\label{eq:nomreg}
  \left\{
  \ket{\mathrm{EPR}^{(1,2)}},
  \ket{\mathrm{EPR}^{(1,3)}},
  \ket{\mathrm{EPR}^{(2,3)}},
  \ket{\mathrm{GHZ}}
  \right\}
\end{equation}
does not suffice.
All these pure states are inequivalent with respect to asymptotic reducibility, but there are pure states that cannot be reversibly generated from these ones alone~\cite{Acin}.
So again, we see that there are \emph{inequivalent kinds of entanglement}.
Because we allowed for asymptotic manipulations, the present inequivalence is even more severe than the one encountered in the last section.

To find general means for constructing
MREGS constitutes one of the challenging open problems
of the field: as long as this question is generally unresolved, the development of a theory of multi-partite entanglement that follows the bipartite example seems unfeasible.
Whereas in the latter case the ``unit'' of entanglement is entirely
unambiguous -- it is the EPR pair~\eqref{eq:EPR} -- there is no substitute for it in
sight for multi-partite systems.

Even if we content ourselves with the building blocks in~\eqref{eq:nomreg}, it is extremely challenging to compute optimal rates for asymptotic conversion.
For example, conversion rates from the GHZ state are directly related to the tensor rank $R_{\min}$ discussed in Section~\ref{subsec:LOCCSLOCC} above.
Consider, e.g., the tripartite state composed out of Bell pairs shared between any pair of subsystems, $\sum_{i,j,k} \ket{i,j}_1 \ot \ket{j,k}_1 \ot \ket{i,k}_1$ (which can be interpreted as a tensor representation of two-by-two matrix multiplication~\cite{landsberg2012tensors}).
The conversion rate from the GHZ state to this tensor is directly related to the computational complexity of matrix multiplication~\cite{chitambar2008tripartite}, a well-known open problem in classical computer science.
Much recent work on multi-partite entanglement transformations has been motivated by this connection~\cite{buhrman2016nondeterministic,vrana2015asymptotic,vrana2016entanglement}.
These intrinsic challenges motivate after all to consider more pragmatic approaches to grasp multi-partite entanglement.

A non-trivial feature of asymptotic LOCC transformations is that multi-partite entangled states can be transformed
into maximally entangled bipartite Bell pairs, fully preserving the entropy of a distinguished party.
More precisely, let $\ket{\psi^{(ABC\dots)}}$ be a state vector of $N+1$ parties. 
Then $\ket\psi$
can be transformed into collections of Bell pairs $\ket{\mathrm{EPR}^{(A_1,B)}},\ket{\mathrm{EPR}^{(A_2,C)}},\dots$, with the property
that $S(\rho^{(A)})$ before is identical with  $S(\rho^{(A_1,A_2,\dots)}) = \sum_{j=1}^N S(\rho^{(A_j)})$ afterwards.
This process -- known as \emph{entanglement combing}\index{entanglement!combing}~\cite{Combing} -- thus
transforms states into a normal form of bipartite entangled states under LOCC.
Note, however, that this transformation is not reversible.

\subsection{Quantifying pure multi-partite entanglement}\label{QuantifyMulti}

Above, we have introduced different classes of entanglement.
It is natural to ask whether there are \emph{entanglement measures}\index{entanglement!measure} that quantify the ``degree of multi-partite entanglement'' found in a state.

There are two approaches to defining entanglement measures: the \emph{axiomatic} and the \emph{operational} one.
In the axiomatic ansatz, one writes down a list of properties one demands from a measure.
The most basic requirement -- referred to as \emph{entanglement monotonicity}\index{entanglement!monotonicity}~\cite{Monotones,Uniqueness} -- is that an entanglement measure must not increase on average under LOCC operations.
More precisely, one requires that
\begin{equation}\label{Monotone}
	E(\rho)\geq \sum_{j=1} p_j E(\rho_j)
\end{equation}
holds true in LOCC protocols in which the state $\rho_j$ is prepared with probability $p_j$ (the label $j$ refers to outcomes of local measurements performed in the course of the protocol).
Convexity is often also taken as a desirable feature~\cite{Monotones,Uniqueness}, even though there are meaningful entanglement measures that are not convex~\cite{NotConvex}.%
\footnote{The term \emph{entanglement measure}\index{entanglement measure} is used somewhat ambiguously in the literature. Some authors distinguish entanglement monotones that satisfy Eq.~\eqref{Monotone} from entanglement measures that merely satisfy $E(\rho)\geq E( \sum_{j=1} p_j \rho_j)$.}
In the multi-partite case, these axioms are rarely strong enough to single out a unique function.
Thus, one must use a certain amount of subjectivity to choose a ``natural'' measure that complies with the axiomatic constraints. Often, ``mathematical simplicity'' is used as a subjective criterion (see below for examples).
The operational approach quantifies the usefulness of a state for a certain protocol that requires entanglement.
Examples would be ``the number of bits of secret key that can be extracted per copy of the state'' in a quantum key distribution scheme, or the ``achievable fidelity'' in a state teleportation protocol.
Here, too, the numbers one obtains strongly depend on a subjective choice: namely which application one has in mind.

Again, let us compare the situation to the bipartite case.
As explained in Section~\ref{sec:asymptotic}, if one allows for protocols operating on asymptotically many copies of a state, the operational ansatz singles out the entanglement entropy as the \emph{unique} pure state measure.
Other pure-state measures are usually only employed if they are easier to treat analytically (such as the \emph{concurrence} introduced below, which can be interpreted as a determinant and inherits some of the simple transformation they enjoy~\cite{verstraete2001local}) or numerically (e.g., the integral R\'enyi entropies of entanglement, which can be estimated using quantum Monte Carlo methods~\cite{broecker2015entanglement}).

Unfortunately, the lack of an explicit multi-partite MREGS means that there is no canonic choice of quantifying entanglement in the general case.
Many measures have been proposed, but none of them is clearly privileged over the others.
(Some aspects of multi-partite entanglement can also be captured by the probability density function of bipartite entanglement~\cite{PhysRevA.74.042331}, such that bipartite properties are inherited in the multi-partite context.)
Here, we briefly describe some of these measures.
For more exhaustive treatments, the reader should consult one of the review articles on the
subject~\cite{HorodeckiEntanglement,amico2008entanglement,plenio2007introduction}.

The \emph{geometric measure of entanglement}\index{geometric measure of entanglement}~\cite{Geometric}
makes use of a geometric distance to the set of product states:
\begin{equation*}
  E_{\text{Geometric}}(\psi) = \min \bigl\lVert \ket\psi\bra\psi  -\sigma \bigr\rVert_2,
\end{equation*}
where $\lVert\cdot\rVert_2$ is the Hilbert-Schmidt norm, and the minimum is taken over all product states $\sigma$.
The construction is very natural: Given that we have already defined the class of unentangled states, maybe the most obvious way of turning it into a quantitative measure is to take the distance of a given state to the unentangled ones.
While it seems not clear whether the geometric measure is \emph{operational} in the sense that its value quantifies the performance of some quantum protocol, several applications are known.
For example, the geometric measure has been used to witness signatures of topological phase transitions~\cite{orus2014topological}.
Also, large values of the geometric measure mean that local measurements will produce highly random results.
This in turn can be used to show that states with a large amount of geometric entanglement are not suitable for certain protocols that are rely on local measurements, most notably \emph{measurement-based quantum computation}~\cite{gross2009most} (cf.\ Section~\ref{sec:mbqc} below).
Distance measures other than the Hilbert-Schmidt norm can also be used, e.g.\ the \emph{relative entropy}\index{entropy!relative}~\cite{RelEntMulti}.

The \emph{Schmidt measure}\index{Schmidt!measure}
$E_S(\psi) = \log R_{\min}(\psi)$
is the logarithm of the minimal number of terms in a product decomposition, as in Eq.~\eqref{schmidtMeasure}.
It is known to be an entanglement monotone~\cite{Schmidt,Graphs,PhD}.
In the bipartite case, this
measure reduces to the \emph{Schmidt rank}\index{Schmidt!rank},
i.e., the rank of either reduced density matrix.
As indicated in the discussion of the GHZ-state, the Schmidt measure can discontinuously increase and decrease.
In contrast, the bipartite Schmidt rank is more benign: around every state there is a neighborhood in which the Schmidt rank does not decrease.
This added instability makes the multi-partite Schmidt measure very challenging to compute numerically.
One can define a  ``smoothed'' version of the Schmidt measure based on a concept -- \emph{border rank} -- from algebraic geometry~\cite{landsberg2012tensors}.
In principle, states of given border rank can be identified as the set of common zeroes of a number of multivariate polynomials.
However, these polynomials are not explicitly known except for a few special cases in low dimensions~\cite{landsberg2012tensors}.

Recall that we have discussed the use of invariants -- e.g.\ the determinant and the hyper\-determinant -- for describing entanglement classes.
It turns out that such functions can be used to construct quantitative entanglement monotones.
Indeed, assume that $f$ is a function of state vectors that is invariant under local $\operatorname{SL}$ operations.
If $f$ is homogeneous of degree two, $f(\lambda\psi) = \lambda^2 f(\psi)$, then $\lvert f(\psi) \rvert$ is an entanglement monotone~\cite{verstraete_normal_2003,miyake_classification_2003}.
This result can be generalized in various ways.
For example, if $f$ is an invariant function of state vectors on $N$ qubits, and if $f$ is homogeneous of degree $d$, then $\lvert f(\psi)\rvert$ is an entanglement monotone if and only if $d\leq 4$~\cite{eltschka2012multipartite}.
This construction is a rich source of entanglement measures.
In the bipartite case, the function $C(\psi) = 2\lvert\det T\rvert$ defined in terms of the coefficient matrix~\eqref{eq:coeff matrix} is such an example.
It is known as the \emph{concurrence}~\cite{hill1997entanglement,wootters1998entanglement}.
Likewise, the hyperdeterminant~\eqref{eq:hyperdet} gives rise to a 3-qubit entanglement measure $\tau_3(\psi)=\lvert\mathrm{Det}(\psi)\rvert$, known as the \emph{3-tangle}\index{3-tangle}~\cite{Tangle,
osterloh2005constructing}
The 3-tangle identifies the GHZ-state as more entangled than the W-state.
It has also been linked to the phenomenon of the \emph{monogamy of entanglement}~\cite{Tangle}.

\subsection{Classifying mixed state entanglement}
\label{mpeMixedClassification}

The program pursued in the preceding section can also be applied to mixed states~\cite{LindenNonLocalDensities,GrasslInvariants}:
One can classify mixed states according to equivalence under various notions of ``local operations'', both for a single or asymptotically many copies, devise quantitative measures, and so forth.
For all these tasks, the situation for pure states is that the bipartite theory is simple, while the multi-partite case quickly becomes complicated.
This changes for mixed states: Here, already the bipartite problems are typically hard!
Even the most elementary question we started this chapter with: ``When can a state be prepared using LOCC?'' is trivial if the state is pure (the answer is ``if and only if it factorizes as in Eq.~\eqref{eqn:factor}''), but NP-hard for bipartite mixed states~\cite{Gurvits,gharibian2010strong}.%
\footnote{One can nevertheless construct \emph{hierarchies} of sufficient criteria for a state being, say, entangled, that are efficiently decidable at each level.
This is possible in a way that every entangled state is necessarily detected in some step of the hierarchy~\cite{Hierarchy,Doh}.
One route towards finding such criteria is to cast the problem into a polynomially constrained optimization problem and relaxing that problem to a hierarchy of efficiently decidable semidefinite programs.
For alternative algorithms for deciding multi-partite entanglement, see Refs.~\cite{Doh,Fer,Bruss}.
}
In light of these difficulties, we will content ourselves with describing only a few entanglement classes for mixed states (this section) and describe a practical method for detecting such classes experimentally in Section~\ref{subsec:detecting} below.

The probably simplest classification of entanglement for mixed states is based on the notion of \emph{separability}~\cite{werner1989quantum,DuerMultiQubitClassification}.
We define an $N$-partite mixed state as unentangled or \emph{fully separable} if it is of the form
\begin{equation}\label{eqn:seperable}
	\rho = \sum_{i} p_i \, \rho^{(1)}_i\otimes\dots\otimes\rho^{(N)}_i,
\end{equation}
for some set of local density matrices $\rho_i^{(j)}$ and a probability distribution $p$.
A mixed state that is not fully separable is \emph{entangled}.
In contrast to unentangled pure states (defined in Eq.~\eqref{eqn:factor}), not every unentangled mixed state is a product state
$\rho^{(1)}\otimes\dots\otimes\rho^{(N)}$ (but every product state is fully separable).
For example, the state
\begin{equation*}
	\rho =
	\frac12\left(
	|0\rangle\langle 0| \otimes |0\rangle\langle 0| \otimes |0\rangle\langle 0|
	+
	|1\rangle\langle 1| \otimes |1\rangle\langle 1| \otimes |1\rangle\langle 1|
	\right)
\end{equation*}
is fully separable, but not a product.
Measurements in the standard basis $\{\ket0, \ket1\}$ on three particles in this state will give perfectly correlated outcomes: all are either found in the $\ket0$ state or all are found in the $\ket1$ state.
This is different from the situation for local measurements on pure unentangled state, which never give correlated outcomes.
The reason such states are defined as unentangled is that they can be created using local operations and classical communication.
Indeed, to create a general state of the form~\eqref{eqn:seperable}, the first party could sample the label $i$ with respect to the probability distribution $p$.
They then communicate the classical information $i$ to all parties.
Upon receiving $i$, the $j$-th party will prepare and output $\rho^{(j)}_i$.
Clearly, when average over the choice of $i$, the output of this preparation procedure is the state $\rho$.

One can refine the classification of mixed entangled states in terms of separability properties.
For example, let us arrange the $N$ parts of the multi-partite system in $k\leq N$ groups, i.e.\ choose a \emph{$k$-partition}\index{partition}.
If we now consider each group as a single party, it could be the case that a previously entangled state becomes fully separable with respect to this coarser partition.
We say that two states belong to the same \emph{separability class}\index{separability class} if they are separable with respect to the same partitions.
Clearly, being in the same class in this sense is a necessary condition for being equivalent under any type of local operations.
A state is referred to as \emph{$k$-separable}\index{k-separable@$k$-separable}\index{separable@separable!k@$k$}, if it is fully separable considered as a state on some $k$-partition.
In this way, we obtain a hierarchy, where $k$-separable classes are considered to be more entangled than the $l$-separable ones for $k<l$.
States that are not separable with respect to any non-trivial partition are called \emph{fully inseparable}\index{inseparable!fully}.

The number of all partitions of a composite system grows exorbitantly fast with the number $N$ of its constituents.
One is naturally tempted to reduce the complexity by identifying \emph{redundancies} in this classification.
After all, once it is established that a state is fully separable, there is no need to consider any further splits.
While such redundancies certainly exist, pinpointing them turns out to be subtle and indeed gives rise to one of the more peculiar results in quantum information theory, as will be exemplified by means of our standard example, the three-qubit system.

The five possible partitions of three systems (1$\mid$2$\mid$3, 1$\mid$23, 2$\mid$13, 3$\mid$12, 123) have already been identified in Section~\ref{subsec:LOCCSLOCC}.
It is a counter-intuitive fact that there are mixed states that are separable with respect to any bipartition, but which are not fully separable~\cite{BennettBiPartiteSeparable}. An analogous phenomenon does not exist for pure states. Specifically, there exist \emph{biseparable}\index{biseparable} (i.e., 2-separable) states of the following kind~\cite{DuerMultiQubitClassification}:
\begin{itemize}
\item \emph{1-qubit biseparable states}, which are separable for 1$\mid$23 but not for 2$\mid$13 nor 3$\mid$12,
\item \emph{2-qubit biseparable states}, which are separable for 1$\mid$23 and 2$\mid$13, but not for 3$\mid$12,
\item \emph{3-qubit biseparable states}, which are separable with respect to any bipartition but nevertheless not fully separable.
\end{itemize}
Together with the fully inseparable states and the fully separable ones, the above classes constitute a complete classification of mixed three
qubit states modulo system permutations~\cite{DuerMixedClassification}.

We end this subsection with a refinement of the class of fully inseparable states that will play a role in the following subsection~\cite{AcinThreeQubits}.
In this paragraph, the fully separable states are denoted by $S$, the convex hull of the biseparable ones by $B$, and lastly, the set of all mixed states including the fully inseparable ones by $F$.
Clearly, $S \subset B \subset F$ is a hierarchy of convex sets.
Now recall that we had identified two different classes of genuinely 3-partite entangled pure states of three qubits in Section~\ref{subsec:LOCCSLOCC}.
We saw that we could approximate a W state up to arbitrary precision by states of GHZ class.
It is not hard to see that biseparable pure states can in turn be approximated by states of W class.
We thus define $W$ to be the set of mixed states that can be decomposed as a convex combination of biseparable ones and W-class states.
This means that $\rho$ is an element of $W$ if three parties can prepare it using local operations, classical communication, and a supply of pure biseparable and W-class entangled states.
Finally, we label the set of all mixed states by $GHZ$.
This leaves us with a finer hierarchy of convex sets
\begin{equation}\label{eq:slocc hierarchy}
  S \subset B \subset W \subset GHZ
\end{equation}
that stratify the space of three-qubit mixed states~\cite{AcinThreeSchmidt1}.
Our preceding considerations showed that a generic three-qubit pure state is of GHZ class.
Hence all other classes of pure states form a subset of measure zero among all pure states.
That notwithstanding, it is easy to see that all three convex sets $S$, $B\setminus S$ and $W \setminus B$ are of nonzero volume in the set of mixed states~\cite{AcinThreeQubits}.
Thus it is meaningful to ask which level of the hierarchy~\eqref{eq:slocc hierarchy} a given mixed quantum state is contained in.

\subsection{Detecting mixed state entanglement}
\label{subsec:detecting}

One way of experimentally detecting multi-partite entanglement is to perform a complete quantum state tomography, and to see whether the resulting estimated state is consistent with an entangled state.
This can be a difficult procedure for two reasons.
First, quantum state tomography amounts to learning the exponentially parameters that describe the quantum state prepared in the experiment, which depending on the number of particles can already be prohibitively costly.
Second, even once we have obtained a complete description of the quantum state, deciding whether the state is in of the separability classes can be a computationally hard problem (even for bipartite mixed states, as we discussed above).

It may be therefore be desirable to detect entanglement without the need to acquire full knowledge of the quantum state (compare the discussion at the end of Section~\ref{subsec:LOCCSLOCC}).
One useful approach is based on the notion of an \emph{entanglement witnesses}\index{entanglement!witness}.
An entanglement witness $A$ is an observable that is guaranteed to have a positive expectation value on the set $S$ of all separable states.
So whenever the measurement of $A$ on some quantum state $\rho$ gives a negative result, one can be certain that $\rho$ contains some entanglement.
It is, however, important to keep in mind that witnesses deliver only \emph{sufficient} conditions.
That is, in addition to $S$ there might be other, entangled states that have a positive expectation value with respect to $A$.

We will now take a more systematic look at this technique and, at the same time, generalize it from $S$ to any compact convex set $C$ in the space of mixed state -- such as the convex sets in the hierarchy~\eqref{eq:slocc hierarchy}!
To that end, we note that the set of quantum states $\sigma$ that satisfy the equation $\tr (\sigma A)=0$ for some observable $A$ forms a \emph{hyperplane}, which partitions the set of states into two half-spaces.
If $C$ is a compact convex set, we can always find a hyperplane such that $C$ is contained in one of these half-spaces, say, $\tr(A\sigma)\geq0$ for all $\sigma\in C$.
Thus, if $\rho$ is a state such that the expectation value of $A$ is negative, $\tr(\rho A)<0$, then, necessarily, $\rho\not\in C$.
It is in this way that entanglement witnesses witness entanglement (more generally, non-membership in some convex set $C$).

Witnesses can be constructed for all of the convex sets that appeared in the classification of the previous subsection.
For example, a \emph{GHZ witness} is an operator that detects certain states that are not of W-type.
It is not difficult to see that
\[ A_{\mathrm{GHZ}} = \frac{3}{4} \id - \ket{\mathrm{GHZ}}\bra{\mathrm{GHZ}} \]
is a GHZ witness: We have that $\langle \mathrm{GHZ} | \rho | \mathrm{GHZ} \rangle\leq 3/4$ and hence that $\tr [A_{\mathrm{GHZ}} \rho]\geq 0$ for any state in the W class.
On the other hand, $\rho = \ket{\mathrm{GHZ}}\bra{\mathrm{GHZ}}$ is a state that will be detected as not being of W-type, since $\tr[A_{\mathrm{GHZ}} \rho] = -1/4$.
More generally, GHZ witnesses can be constructed as $A_{\mathrm{GHZ}} = Q-\varepsilon \id$ with an appropriate $\varepsilon>0$, where $Q\geq 0$ is a matrix that does not have any W-type state in its kernel.
Similarly, one possible \emph{W witness} is given by
\begin{equation}\label{wWitness}
  A_{\mathrm{W}}= \frac23 \id -  \ket{\mathrm{W}}\bra{\mathrm{W}}.
\end{equation}
Expectation values of witness operators can be obtained from local measurements by using appropriate local decompositions~\cite{Guehne}, in the same way as one can choose a basis consisting of product matrices when performing a tomographic measurement.
The detection of multi-partite entanglement using witness operators has already been experimentally realized~\cite{WeinDetect}. Indeed, one of the witness operators that was estimated in this experiment was of the form given in~\eqref{wWitness}.

\section{Important classes of multi-partite states}\label{sec:importantstates}

A pure $N$-qubit state is specified by $2^{N}$ complex amplitudes.
Of course, no-one can make sense of, say $\simeq 1000$ complex numbers that specify a 10-qubit state.
What is more: Physical preparation procedures that require more than a polynomial number of parameters to be described are impractical to implement.
As the number of particles grows, it follows that ``most'' states cannot be realistically prepared and will thus never occur in natural nor in engineered quantum systems.
Therefore, multi-partite entanglement theory is relevant only in as much as there are many-body states that \emph{i}) exhibit interesting features, \emph{ii}) allow for a description in terms of polynomially many parameters, and that \emph{iii}) can be efficiently created. Fortunately, such family of states are known.
The arguably most prominent examples are \emph{tensor network states}, relevant e.g.\ in condensed matter theory and, more recently, in high energy theory, and
explained in Section~\ref{sec:manybody},
\emph{stabilizer states}, described in Section~\ref{sec:stabilizers}, and
\emph{bosonic and fermionic Gaussian states}, mentioned in Section~\ref{sec:gaussian}.

\subsection{Matrix-product states and tensor networks}\label{sec:manybody}

Quantum systems with many degrees of freedom are ubiquitous in nature, particulary in the context of condensed matter theory.
It is hence not surprising that important classes of states, such as ground states of local Hamiltonians, are multi-partite entangled states.
This viewpoint goes beyond a mere curiosity and provides a relevant perspective when describing properties of quantum many-body systems.
Recent years have seen an enormous increase in interest at the intersection of quantum information and condensed matter theory that stems from the insight that notions of entanglement are crucial in the understanding of quantum phases of matter.
It goes beyond the scope of this chapter to elaborate in detail on this connection, but we will briefly describe a number of important insights.

To start with, the complexity of natural multi-partite states, such as ground states of local Hamiltonians in a gapped phase, is significantly reduced by the insight that they often satisfy what is called an \emph{area law}\index{area law} for the entanglement entropy.
To make this more precise, recall that a \emph{local Hamiltonian}\index{focal Hamiltonian}\index{Hamiltonian!local} is a Hamiltonian of the form $H=\sum_j h_j$, where each $h_j$ is supported only on at most $k$ (usually $k=2$) subsystems, describing a short-ranged interaction.
\emph{Gapped phase}\index{gapped phase} refers to the fact that the lowest energy level, the smallest eigenvalue of $H$, is separated by a gap $\Delta>0$ that is
uniform in the thermodynamic limit of letting the number of particles $N\rightarrow \infty$.
Such gapped phases describe realistic condensed matter systems away from quantum phase transitions.
Finally, an \emph{area law}\index{area law} states that, for any subsystem $A\subset \{1,\dots, N\}$, the entanglement entropy $S(\rho^{(A)})$ of the reduced state associated with $A$ scales as
\begin{equation*}
	S(\rho^{(A)})= O(\lvert\partial A\rvert),
\end{equation*}
where $\partial A$ is the boundary of $A$ and $\lvert\partial A\rvert$ its area.
In one-dimensional systems, $A$ is a union of intervals and this area refers to the number of endpoints, which means that $S(\rho^{(A)})$ is upper bounded by a constant \emph{independent} of the system size $N$.
This feature is remarkable and it points towards the way ground states of natural physical models are non-generic.
Haar random states, e.g., would with overwhelming probability lead to states for which $S(\rho^{(A)})$ scales close to extensively.
Ground states of gapped models deviate from this generic behavior and are much less entangled than they could potentially be.
Such area laws have been proven for one-dimensional systems~\cite{OneD} as well as in arbitrary dimensions for non-interacting bosonic and fermionic models~\cite{PhysRevA.73.012309,AreaReview}.
This scaling of the entanglement entropy offers profound insight into the structure of entanglement in quantum many-body systems (that can be extended to the study of non-leading corrections to $S(\rho^{(A)})$ that diagnose \emph{topological order}\index{topological order}~\cite{kitaev2006topological,levin2006detecting}).
More concretely, it suggests that one can largely parametrize the subset of Hilbert space ${\cal H}$ that is occupied by ground states of gapped local models.
This hope is indeed largely fulfilled by so-called \emph{tensor network states}\index{tensor network state}~\cite{Orus-AnnPhys-2014}.

This picture is particularly clear for one-dimensional systems. Here, \emph{matrix product states}\index{matrix product state}~\cite{raey,MPSRev} provably provide a good approximation to ground states of gapped local Hamiltonians in terms of polynomially many parameters only.
Suppose that each site $j=1\,\dots, N$ is of local dimension $d$ and equipped with a tensor of order three, which can equivalently be seen as a collection of
matrices $M^{(j)}[1] ,\dots, M^{(j)}[d]\in \cc^{D\times D}$; $d$ is called the \emph{physical dimension} and $D$ the \emph{bond dimension}.
This data defines pure quantum state of $N$ $d$-dimensional particles -- a matrix product state:
\begin{equation*}
	\ket\psi = \sum_{x_1,\dots,x_N=1}^d \tr\left( M^{(1)}[x_1] M^{(2)}[x_2]\dots   M^{(N)}[x_N]\right) \ket{x_1}\otimes\ket{x_2}\otimes\dots\otimes\ket{x_N}.
\end{equation*}
This is a huge dimensional reduction:
Instead of having to deal with $\Theta(d^N)$ many parameters, $\Theta(N d D^2)$ many are sufficient.
Not only do such states satisfy an area law for the entanglement entropy.
More importantly, the converse is also true:
Each quantum state that satisfies an area law for (R\'enyi) entanglement entropies can be approximated in trace norm by a matrix product state of polynomial bond dimension $D$~\cite{SchuchApprox}.

The significance of these insights can hardly be overestimated:
They are at the heart of the functioning of the various variants of the \emph{density matrix renormalization group} approach~\cite{DMRGWhite92,MPSRev}, which solves strongly correlated one-dimensional models essentially to machine precision.
In additional to serving as variational states in powerful numerical methods, they can be used as a mathematical tool to capture properties of interacting quantum many-body systems.
For example, the problem of classifying the quantum phases of matter in one-dimensional systems in the presence of symmetries has been rigorously solved in this way~\cite{1010.3732,PhysRevB.86.125441}.
It should be clear that matrix product states can exhibit multi-partite entanglement for pure quantum states in any of the senses described above.

More generally, one can define \emph{tensor network states}\index{tensor network state} by placing tensors at the vertices of an arbitrary graph.
Each tensors carries an index of the physical dimension $d$ and further indices of the bond dimensions $D$, corresponding to the edges incident to the vertex.
These tensors are then contracted (that is, summed up) according to the edges of the graph.
Matrix product states are tensor network states corresponding to a linear graph (with edges $i-i+1$).
In higher dimensions, \emph{projected entangled pair states}~\cite{Orus-AnnPhys-2014} are based on cubic lattices. 
Again, such states satisfy area laws by construction and serve as good variational states for ground states of local Hamiltonians~\cite{Orus-AnnPhys-2014} (although this has not been mathematically proved in generality).
They also serve as resources for measurement-based quantum computing ~\cite{Oneway,Darmawan,gross2007novel}, discussed in Section~\ref{sec:mbqc} below.
The \emph{multi-scale entanglement renormalization ansatz (MERA)}\index{multi-scale entanglement renormalization ansatz}\index{MERA}~\cite{vidal2007entanglement,vidal2008class} describes ground states at quantum critical points and provides a new perspective on renormalization.
In high energy physics, tensor network models based on the MERA have featured in quantum information theoretic approaches to the holographic duality~\cite{swingle2012entanglement,swingle2012constructing,qi2013exact,pastawski2015holographic,yang2015bidirectional,hayden2016holographic}.

\subsection{Stabilizer states}\label{sec:stabilizers}

Stabilizer states and their generalizations form the basis of the theory of quantum error correcting codes~\cite{gottesman96stabilizer,nielsen2000quantum}.
They are also used for measurement-based quantum computation~\cite{Oneway}, violate many-party Bell inequalities~\cite{guhne2005bell}, can exhibit topological order~\cite{kitaev2003fault,zeng2015quantum}, and emulate, in a precise sense, some features of ``generic states''~\cite{zhu2016clifford,nezami2016multipartite,webb_clifford_2015}.
Stabilizer states can be defined as the unique common $(+1)$-eigenvector of subsets of \emph{Pauli operators}.
We will not describe the general theory here (see Ref.~\cite{nielsen2000quantum}), but we will try to convey the flavor of it.
To this end, recall the single-qubit Pauli operators
\begin{equation*}
	\mathbb{I}=\begin{pmatrix}1&0\\0&1\end{pmatrix},
	\quad
	X=\begin{pmatrix}0&1\\1&0\end{pmatrix},
	\quad
	Y=\begin{pmatrix}0&-i\\i&0\end{pmatrix},
	\quad
  Z=\begin{pmatrix}1&0\\0&-1\end{pmatrix}.
\end{equation*}
To give a first example, it is easy to see that the state vector
\begin{equation*}
	\ket{\mathrm{GHZ}} = \frac 1 {\sqrt 2} \left( \ket{0,0,0}+ \ket{1,1,1} \right)
\end{equation*}
is an eigenvector of the three-qubit Pauli operators
\begin{equation}\label{eqn:ghz_stab}
	Z_1\otimes Z_2\otimes \id_3, \qquad
	\id_1\otimes Z_2\otimes Z_3, \qquad
	X_1\otimes X_2\otimes X_3
\end{equation}
to the eigenvalue $+1$.
In fact, it is unique state with that property.
So instead of explicitly writing down the expansion coefficients of the GHZ state vector with respect to a basis, we can specify it implicitly as the vector \emph{stabilized} by the three Pauli operators in Eq.~\eqref{eqn:ghz_stab}.
The advantage of this approach becomes apparent only as the number of qubits grows.
Pauli operators on $N$ qubits are tensor products of $N$ single-qubit Pauli operators, and an $N$-qubit stabilizer state can be uniquely defined as the common $(+1)$-eigenvector of $N$ Pauli operators.
Thus, the complexity of specifying stabilizer states grows only as $O(N^2)$ -- much more favorably than the exponential scaling required when naively writing out expansion coefficients.

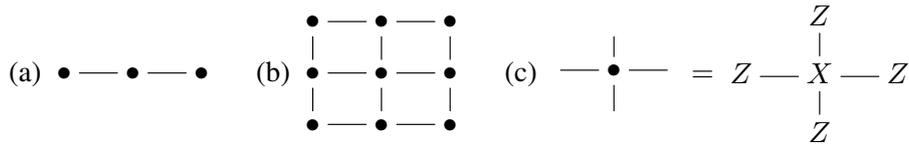
\begin{figure}
	\begin{center}
		\begin{tabular}{ccc}
		(a)
		$
			\begin{xy}
			*!C\xybox{\xymatrix@C=5mm@R=3mm{
\bullet \ar@{-}[r]& \bullet \ar@{-}[r]& \bullet
			}}
			\end{xy}
		$
		\quad
		&

		(b)
		$
			\begin{xy}
			*!C\xybox{\xymatrix@C=5mm@R=3mm{
\bullet \ar@{-}[r]& \bullet \ar@{-}[r]& \bullet \\
\bullet \ar@{-}[u]\ar@{-}[r]& \bullet \ar@{-}[u]\ar@{-}[r]& \bullet \ar@{-}[u] \\
\bullet \ar@{-}[r]\ar@{-}[u]& \bullet\ar@{-}[r] \ar@{-}[u]& \bullet \ar@{-}[u]
			}}
			\end{xy}
		$
		\quad
		&

		(c)
		$
			\begin{xy}
			*!C\xybox{\xymatrix@C=5mm@R=3mm{
	&  \\
 \ar@{-}[r]& \bullet \ar@{-}[u]\ar@{-}[r]& \\
	& \ar@{-}[u]
			}}
			\end{xy}
			=
			\begin{xy}
			*!C\xybox{\xymatrix@C=5mm@R=3mm{
	& Z \\
Z \ar@{-}[r]& X \ar@{-}[u]\ar@{-}[r]& Z  \\
	& Z  \ar@{-}[u]
			}}
			\end{xy}
		$
	\end{tabular}
\end{center}

\caption{\label{fig:graph_state}
	Quantum states associated with graphs.
	(a)~Defines the \emph{linear cluster state} of length 3.
	(b)~Graph states corresponding by a two-dimensional lattice are of particular interest, e.g.\ in measurement-based quantum computation.
	(c)~Graph states can be defined via a simple preparation procedure: Vertices correspond to qubits initially in $\ket+$; edges denote the application of a controlled-Z gate. Graph states are stabilizer states: Every vertex defines an element of the stabilizer group as follows. Associate a
	Pauli-$X$ matrix acting on the qubit that belongs to the given vertex, with Pauli-$Z$ matrices acting on every qubit in its (graph-theoretical) neighborhood.
}
\end{figure}

There is a subset of all stabilizer states, called \emph{graph states}~\cite{Graphs}.
They allow for a particularly intuitive and physical description.
Every stabilizer state can be brought into the form of a graph state using only local unitaries~\cite{Schlinge} -- so not much generality is lost by focusing on this special case.
A graph state is defined by a \emph{graph} with one vertex for every qubit (cf.~Fig.~\ref{fig:graph_state}).
To obtain the associated quantum state, we first prepare every qubit in the
$\ket + = (\ket 0 + \ket 1)/\sqrt{2}$-state vector and then apply a \emph{controlled-$Z$}
\index{Controlled-$Z$-gate}
gate
\begin{equation*}
	CZ: \ket{x_1, x_2} \mapsto (-1)^{x_1 x_2} \ket{x_1, x_2}
\end{equation*}
between any two qubits that are joined by an edge in the graph.
Because controlled-Z gates acting on different pairs of qubits all commute with each other, the order in which this process is carried out is immaterial.
As an example, it is instructive to verify that the graph in Fig.~\ref{fig:graph_state}~(a) defines the state vector
\begin{align*}
		&\quad\ket{\text{Cluster}_3} \\
		&=
		2^{-3/2}
		\bigl(
			\ket{0,0,0}
			+\ket{0,0,1}
			+\ket{0,1,0}
			-\ket{0,1,1}
			+\ket{1,0,0}
			+\ket{1,0,1}
			-\ket{1,1,0}
			+\ket{1,1,1}
		\bigl) \\
		&=
		2^{-1/2}
		\bigl(
			\ket{+,0,+}
			+\ket{-,1,-}
		\bigl),
\end{align*}
which is known as the \emph{three-qubit cluster state}\index{cluster state!three-qubit}.
Here, we have used the abbreviation $\ket\pm = 2^{-1/2} (\ket0\pm\ket1)$.
A unitary basis change $\ket+\mapsto\ket0, \ket-\mapsto\ket1$ on the first and the third qubit maps the three-qubit cluster state to the GHZ state we have encountered before.

One of the most heavily studied graph states is the one that corresponds to a two-dimensional $n\times n$-lattice (Fig.~\ref{fig:graph_state}).
It is known as the \emph{two-dimensional cluster state}\index{cluster state!two-dimensional}.
Unlike the three-qubit linear cluster encountered above, it would be completely impractical to write out the expansion coefficients of these states for larger values of $n$.
The relevance of the two-dimensional cluster state comes from the fact that simulating the correlations between local measurements on it is provably as difficult as predicting the outcome of any quantum computation~\cite{Oneway}, as briefly discussed in Section~\ref{sec:mbqc} below.

Thanks to their algebraic structure, the entanglement structure of stabilizer states is much better understood than for general quantum states.
In particular, any tripartite pure stabilizer state admits a simple normal form:
It can by local (Clifford) unitaries be converted into a tensor product of fully separable product states, bipartite Bell pairs and tripartite GHZ states~\cite{bravyi2006ghz}.
Thus~\eqref{eq:nomreg} constitutes a complete set of building blocks of tripartite entanglement for stabilizer states, and there is a unique unit of \emph{genuinely} tripartite entanglement, the GHZ state.
The amount of Bell pairs and GHZ states contained in a given stabilizer state can be readily obtained from certain invariants~\cite{bravyi2006ghz,smith2006typical,nezami2016multipartite}.
For more than three subsystems, however, the situation is again less clear.

\subsection{Bosonic and fermionic Gaussian states}\label{sec:gaussian}

Another family of quantum many-body states that can be efficiently described are the classes of \emph{bosonic and fermionic Gaussian states}\index{Gaussian state}\index{Gaussian state!of bosons}\index{Gaussian state!of fermions}\index{bosons!Gaussian state}\index{fermions!Gaussian state}.
They both arise naturally in the context of quantum many-body models in condensed matter physics, but their bosonic variant is also highly useful in quantum optics when it comes to describing systems constituted of several quantum modes of light.

Bosonic and fermionic Gaussian systems consisting of $N$ modes can be described in a very similar fashion, and our subsequent brief description will stress this aspect.
For comprehensive reviews, we refer to Refs.~\cite{GaussianIntro,RevModPhys.84.621,AreaReview}.
Bosonic systems are equipped with $2N$ Hermitian canonical coordinates corresponding to position
$x_1,\dots, x_N$ and momentum
$p_1,\dots, p_N$. Once collected in a vector $R=(x_1,\dots, x_N,p_1,\dots, p_N)$, the
familiar \emph{canonical commutation relations} take the form
\begin{equation*}
	[R_j, R_k] =  i \sigma_{j,k},\,\,\,\, \sigma=
\begin{pmatrix}
0&\id \\ -\id&0
\end{pmatrix}.
\end{equation*}
For fermionic systems, one can similarly
define \emph{Majorana fermions} $c_1,\dots, c_{2N}$, Hermitian operators that satisfy
	$\{c_j,c_k\}=\delta_{j,k}$,
taking a similar role as canonical coordinates do.
As for bosons, they are linear combinations of fermionic annihilation and creation operators.

There are several equivalent ways of defining Gaussian states: One way is to
define Gaussian states as the Gibbs states of Hamiltonians that are quadratic polynomials
in the canonical coordinates or the Majorana operators.
The central object in the study of Gaussian states is the covariance matrix, a statement that applies to both
bosonic and fermionic systems. Fermionic Gaussian states are actually completely defined by
their covariance matrix, as the first moments (i.e., expectation values) of Majorana operators necessarily vanish due to the
parity of fermion number superselection rule. Bosonic Gaussian state are specified by their
covariance matrix together with the first moments of the canonical coordinates, but also here we will
focus on the covariance matrix alone.
Specifically, for bosonic systems, the \emph{covariance matrix}\index{covariance matrix}  is a real symmetric $2N\times 2N$
matrix $\gamma_B= \gamma_B^T$ with entries $(\gamma_B)_{j,k} = \tr((R_j R_k +R_k R_j)\rho)$.
It satisfies $\gamma_B+ i \sigma \geq 0$, reflecting the Heisenberg uncertainty relation.
Under mode transformations, covariance matrices transform as $\gamma\mapsto S\gamma S^T $.
Here, $S$ is a matrix that satisfies $S\sigma S^T=\sigma$, which implies that the mode transformation preserves the canonical commutation relations.
Such matrices form a group, the real symplectic group $Sp(2N, \mathbb{R})$.
A helpful tool is the \emph{normal mode decomposition}\index{normal mode decomposition}, also called \emph{Williamson normal form}\index{Williamson normal form}, which maps a covariance matrix into one that describes $n$ uncorrelated modes,
\begin{equation*}
	S\gamma S^T = \bigoplus_{j=1}^n
  \begin{pmatrix}
  D_j &0\\0&D_j
  \end{pmatrix}.
\end{equation*}
For any covariance matrix (in fact, for any strictly positive operator), a suitable $S\in Sp(2N, {\mathbb{R}})$ can always be found.
The values $\{D_j\}$ are called the \emph{symplectic eigenvalues}\index{eigenvalue!symplectic}.
For a pure Gaussian state, they would take the value $D_j=1$ for all $j$.
This is a most convenient instrument, as in this way one can often decouple correlated problems and thus relate the computation of unitarily invariant quantities, such as of the von Neumann entropy, to expressions involving single modes only.
An example of a family of covariance matrices of a pure Gaussian states of three modes is
\begin{equation*}
\gamma=
\begin{pmatrix}
a & b & b\\
b &a & b\\
b & b & a
\end{pmatrix}
\oplus
\begin{pmatrix}
a & c & c\\
c &a & c\\
c & c & a
\end{pmatrix},
\end{equation*}
for $a>1$ and $c,b=-(1 \pm a^2 +\sqrt{1 - 10 a^2 + 9 a^4})/(4 a)$.
For any value of $a>1$, this is the covariance matrix of a genuinely three-party entangled state that in some ways takes the role of the GHZ state for three qubits. The tri-partite entanglement features of such states have been discussed in Refs.~\cite{PhysRevA.73.032345,GaussianQuantumMarginals}.

For fermionic systems, the covariance matrix is again a real $2N\times 2N$ matrix $\gamma_F$, which
is now anti-symmetric, $\gamma_F=-\gamma_F^T$. Its entries are
$(\gamma_B)_{j,k} = i \tr(\rho[c_j,c_k])$.
By construction, it satisfies $\gamma_F^T\gamma_F\leq \id$.

Covariance matrices and the corresponding phase space transformations make up the \emph{phase space formalism} for bosonic and fermionic Gaussian states.
Since these are $(2N\times 2N)$ matrices, this gives rise to an efficient description.
Their importance in quantum optics and condensed matter theory can hardly be overstated.
Most quantities of interest, including bipartite and multi-partite entanglement measures, can be computed directly from the covariance matrices.

\section{Specialized topics}
\label{sec:specialized}

\subsection{Quantum Shannon theory}

An important field in quantum information theory is \emph{quantum Shannon theory}\index{quantum Shannon theory}, which is concerned with the transmission of classical and quantum information via quantum communication channels
(see, e.g., Refs.~\cite{nielsen2000quantum,wilde2011classical}).
For a subsystem $A$, let us write $S(A) =  S(\rho^{(A)})$ for the \emph{von Neumann entropy}\index{entropy!von Neumann} of its reduced density matrix.
Optimal rates for a wide range of communication tasks are given by suitable linear combinations of subsystem entropies.

Since we may always imagine a quantum system to be in an overall pure state, the von Neumann entropy $S(A)$ is the same as the entropy of entanglement between the system $A$ and its environment, as discussed in Section~\ref{sec:asymptotic}. It is easy to see that the entanglement entropy is nonnegative and bounded by the logarithm of the Hilbert space dimension, $0\leq S(A)\leq \log \dim A$.
The entropies of individual subsystems are not independent, but rather constrained by linear entropy inequalities.
For example, the \emph{mutual information}\index{mutual information} $I(A:B) = S(A) + S(B) - S(AB)$ is never negative.
It is intuitive, but much more difficult to show, that the mutual information can never increase when we discard subsystems:
\begin{equation}\label{eq:monotone}
  I(A:B) \leq I(A:BC)
\end{equation}
Equivalently, $S(AB)+S(BC)\geq S(B)+S(ABC)$, which is known as the \emph{strong subadditivity}\index{strong subadditivity} property of the von Neumann entropy~\cite{lieb1973fundamental}.
Strong subadditivity is a fundamental tool in quantum information theory, and there has been much recent progress on deriving \emph{stronger} variants~\cite{winter2012stronger,fawzi2015quantum}.
As a simple but illustrative example, we can use this \emph{multi-partite} entropy inequality to show that it is not possible for a sender Alice to communicate more than $N$ classical bits by sending $N$ qubits alone.
This result is known as \emph{Holevo's theorem}\index{Holevo's theorem}~\cite{holevo1973bounds}.
To start, suppose that Alice carries a classical register $X$.
Depending on the value of $X$, she prepares one of many $N$-qubit states $\rho_x$, and sends it over to Bob.
The joint state of Alice and Bob can be modeled by a mixed state
\[ \rho^{(XB)} = \sum_x p_x \ket x\bra x^{(X)} \ot \rho_x^{(B)}. \]
Bob subsequently carries out a POVM measurement on his system and obtains a classical outcome $Y$.
Such a measurement can always be realized by first applying an isometry $B\to YC$, where $C$ is an auxiliary subsystem, and subsequently discarding $C$.
Using~\eqref{eq:monotone}, we find that $I(X:Y) \leq I(X:YC)$. But $I(X:YC) = I(X:B) \leq N$, since $N = \log \dim B$ is the number of qubits in $B$.
This shows that Bob cannot learn more than $N$ classical bits of information about Alice's register, as quantified by the mutual information $I(X:Y)$.

While \emph{all} constraints on quantum entropies for bipartite (and pure three-party) systems are known,
it is an important open question to decide if there are additional inequalities other than strong subadditivity in the multi-partite case~\cite{pippenger2003inequalities,linden2005new,cadney2012infinitely,christandl2012recoupling,linden2013quantum,gross2013stabilizer}.
Such \emph{non-Shannon information inequalities} are known to exist in classical information theory~\cite{zhang1998characterization,dougherty2006six,matus2007infinitely}.

\subsection{Quantum secret sharing and other multi-party protocols}

Relatedly, multi-partite entangled states serve as resources to a number of important protocols in quantum information theory in which more than two parties come together.
A prominent example of such a \emph{multi-party quantum protocol}\index{quantum protocol!multi-party} is \emph{quantum secret sharing}\index{quantum secret sharing}~\cite{PhysRevA.59.1829,PhysRevLett.83.648}, in which a
message is distribute to several parties in such a way that no subset is able to read the message, but the
entire collection of parties is.
The above-mentioned GHZ states constitute resources for such protocols.
The basic idea is rather natural:
For an $N$-qubit system with parties labeled $\{1,\dots, N\}$, consider the generalized GHZ state $(\ket{0,\dots, 0} + \ket{1,\dots,1})/2^{N/2}$.
It is clear that the reduced state of any proper subset $A\subset \{1,\dots, N\}$ will satisfy
\begin{equation*}
	\rho^{(A)}
  = \frac{1}{2^{\lvert A\rvert}} \left( \ket0\bra0^{\ot A} + \ket1\bra1^{\ot A} \right).
\end{equation*}
But the same will be true for $(\ket{0,\dots,0} - \ket{1,\dots,1}/2^{N/2}$.
Thus the two states can only be distinguished when all $N$ parties come together.
More generally, one speaks of a $(t,N)$ threshold scheme if one can divide a secret into $N$ shares such that $t$ of those shares can be used to reconstruct the secret, while any  $t-1$ or fewer shares reveal no information about the secret at all.
The security of quantum secret sharing has been discussed in Ref.~\cite{SecuritySecretSharing}.
A type of multi-party entangled state that features in the discussion of quantum secret sharing schemes is the one of \emph{absolutely maximally entangled states}\index{absolutely maximally entangled states}, which are characterized by being maximally entangled for all bipartitions of the system~\cite{PhysRevA.77.060304}.
For qubit systems, they exist only for a particular choices of $N$.
Specifically, no absolutely maximally entangled states exist for $N=4, 8$ or $N>8$~\cite{GourAME,RainsCodes}.
The also do not exist for $N=7$~\cite{GuehneAME}.
For suitable local dimension $d>2$, absolutely maximally entangled states exist for all $N$~\cite{PhysRevA.86.052335} (cf.~\cite{ding2016conditional}).
The notion of absolutely maximally entangled states also relates to observations that some aspects of multi-partite entanglement may be captured in terms of the purity of balanced bipartitions, made up of half of the subsystems.
When several bipartitions are considered at the same time, the requirement that the purity be minimal can lead to frustration~\cite{Ugo}.
Absolutely maximally entangled states have to be distinguished from the maximally entangled set of multi-partite quantum states~\cite{KrausMaximallyEntangled} described earlier in Section~\ref{subsec:LOCCSLOCC}.

In addition to secret sharing, a number of other important multi-party quantum protocols have been introduced that directly use multi-partite entangled states as resources.
Many of these schemes have a relationship to cryptography, going beyond key distribution in point-to-point architectures.
Notably, notions of \emph{secure function evaluation}\index{secure function evaluation}~\cite{SecureFunctionEvaluation}
have been introduced, again in a multi-partite setting.
Certain Calderbank-Shor-Steane (CSS) states, which are instances of stabilizer states as described in Section~\ref{sec:stabilizers} above, can be made use of to devise ``prepare and measure'' protocols for quantum cryptography that can be employed in a \emph{conference key agreement}~\cite{PhysRevA.67.012322}, a protocol that allows a number of parties to share a secure conference key.
Relatedly, the \emph{quantum sharing of classical secrets}~\cite{PhysRevA.68.062108} has been proposed.
Protocols such as the above are expected to become particularly prominent once multi-party quantum networks~\cite{QuantumInternet} become available.

The use of multi-partite resources can also give rise to other practical or technological advantages: For example, photonic architectures for measurement-based quantum computing become more efficient once entangled GHZ states are used as a resource~\cite{PhysRevLett.115.020502}, compared to schemes built on bipartite entangled photonic states.

\subsection{Quantum non-locality}
\label{subsec:non-locality}

It is a remarkable fact that quantum mechanics gives rise to correlations that are not compatible with any local hidden variable theory.
Ever since the original \emph{Einstein-Podolski-Rosen (EPR) paradox}\index{EPR paradox}~\cite{einstein1935can}, this has been of much debate and interpretation.
As we will now illustrate, multi-partite quantum states give rise to new and exotic correlations that in some ways sharpen Bell's famous theorem~\cite{bell1964einstein,clauser1969proposed}.

We will illustrate this with an example, phrased in the modern language of \emph{non-local games}\index{non-local games}\index{non-locality}.
In the \emph{Mermin GHZ game}\index{Mermin GHZ game}\index{GHZ game}~\cite{mermin1990extreme}, three players Alice, Bob and Charlie each receive an input bit, $x$, $y$, and $z$, respectively, from the referee, with the promise that $x+y+z$ is even.
There are four such options: either all three bits are zero, or there is a single zero bit and two one bits (there are three such options).
We will assume that the referee sends each such option with equal probability.
The goal of the three players, who are not allowed to communicate after having received their inputs, is to output bits $a$, $b$, and $c$, respectively, such that $a+b+c$ is even in the first case, and odd otherwise. Mathematically, we require that $a+b+c=x\vee y\vee z$ modulo two.

It is easily understood that the GHZ game cannot be won by any classical local strategy.
Indeed, if the three players follow deterministic strategies $a(x)$, $b(y)$, and $c(z)$, respectively, they succeed if and only if
\begin{align*}
  a(0)+b(0)+c(0)=0, \quad a(1)+b(1)+c(0)=1, \\
  a(1)+b(0)+c(1)=1, \quad a(0)+b(1)+c(1)=1.
\end{align*}
If we sum these equations modulo two, we obtain that $0=1$, a contradiction.
Moreover, shared randomness does not help -- the classical winning probability remains $3/4$.

In contrast, if the three players share a GHZ state $(\ket{0,0,0}+\ket{1,1,1})/\sqrt 2$ then they can win this game \emph{with certainty}.
For this, each of the players proceeds as follows:
Depending on whether their input is zero or one, they either measure in the $X$ or in the $Y$ eigenbasis.
If the measurement result is $(-1)^m$, they output $m$.
It is readily verified in a few lines that this strategy is always successful.
We note that GHZ non-locality has been tested experimentally~\cite{pan2000experimental}
For further detail on the thriving field of non-local games we refer to the recent survey~\cite{palazuelos2016survey}.

\subsection{Measurement-based quantum computing}\label{sec:mbqc}

Computers that make use of the laws of quantum mechanics are strongly believed to outperform any classical architecture for certain problems~\cite{nielsen2000quantum}.
According to the widely-used \emph{gate model}\index{gate!model}, a quantum computation proceeds as follows: First a number of qubits are initialized in some product state, e.g.\ $\ket0^{\otimes N}$.
Then a sequence of \emph{quantum gates} is applied.
Quantum gates are unitary operations that act non-trivially only a small number of the $N$ systems, usually
one or two.
Quantum gates are a natural analogue to the \emph{logical gates} that appear both in the mathematical description of the classical \emph{circuit model} of computation, and in the silicon hardware of actual computers.
The algorithm to be performed is encoded in the choice of gates.
In a final step, each of the qubits is measured in some basis.
The measurement outcomes define $N$ bits, which are the result of the quantum computation.

The time evolution generated by the unitary gates is thus the central ingredient to a gate-model quantum computation.
Given this situation, it was a major discovery that there are ways to realize arbitrary quantum algorithms without any unitary evolution at all.
The \emph{measurement-based model of quantum computation}\index{measurement-based!quantum!computation} (MBQC)~\cite{Oneway} only employs local measurements on an entangled many-body quantum state.
These protocols start with a certain \emph{universal resource state}\index{resource!state} on $N$ qubits.
The state does not depend on the specific quantum computation we aim to perform, other than that the size $N$ has to be large enough to support it.
Now assume the algorithm is specified in the form of a sequence of unitary gates borrowed from the gate model.
Ref.~\cite{Oneway} specifies a simple translation rule that maps the gate sequence to a measurement protocol.
The protocol calls for the qubits to be measured one by one.
The respective measurement basis depends on the gate sequence as well as on the previous outcomes.
After all $N$ qubits have thus been measured, a simple classical post-processing algorithm extracts the result of the gate-model computation from the local measurement outcomes.

It is beyond the scope of the present chapter to present the detailed functioning of measure\-ment-based quantum
computing and refer to the original articles. However, the basic functioning can be made plausible by what
is called one-bit teleportation: Take a qubit prepared in the state vector $\ket{\psi}$ and another one in $\ket{+}$,
and subsequently apply a controlled-$Z$ gate as described in Section~\ref{sec:stabilizers}. Then the following identity
will be true,
\begin{equation*}
	(\bra{m}\otimes \id)CZ \ket{\psi}\ket{+} = X_2^m H_2 \ket{\psi}.
\end{equation*}
That is, upon measurement of the first qubit, the second qubit will be in $\ket{\psi}$,
up to an application of a Hadamard gate and a Pauli-$X$ operator applied to the second qubit,
dependent on the specific measurement outcome $m\in \{0,1\}$. By means of this operation,
the state vector has hence be shifted by one site, up to the application of a Pauli operator. Concatenating
such steps, one can show that one can transport an arbitrary state through the entire lattice. What is more,
upon changing the
measurement basis, an actual computation can be done, without ever to have to physically implement
any unitary gate.

The discovery of MBQC has established a novel facet of multi-partite entanglement.
Namely, we can now classify many-body quantum states by their ability (of lack thereof) to boost the computational power of a classical control computer. In this framework,
the resource character of entanglement  for accomplishing computational tasks is most transparent.
Several examples of universal resource states of MBQC are known.
The most prominent one is the cluster state on a square lattice (cf.~Section~\ref{sec:stabilizers} and Fig.~\ref{fig:graph_state}~(b)).
It was a further important insight to see that further resource states can be constructed from tensor network states~\cite{gross2007measurement}.
Since then, more general families of states have been identified~\cite{gross2007novel,brennen2008measurement,miyake2011quantum},
often based on their tensor network description. 

While the quality of ``being a resource state for MBQC'' is a legitimate facet of multi-partite
entanglement in itself, one can ask how it relates to other measures.
High values of several multi-partite entanglement measures -- most prominently the \emph{localizable entanglement}~\cite{Localizable} -- have been linked to universal resource states~\cite{van2006universal}.
Conversely, it has been found that high value of the geometric measure are detrimental for MBQC~\cite{gross2009most}.

\subsection{Metrology}

Multi-partite entanglement does not only facilitate processing or
transmission of information, but also allow for
applications in \emph{metrology}\index{metrology}~\cite{Metrology,Metrology2,WineOld,Frequency}.
We will shortly sketch an idea to enhance the accuracy
of the estimation of frequencies using multi-partite
entangled states. This applies in particular
to frequency standards based on laser-cooled ions, which
can achieve very high accuracies~\cite{Frequency}.
The starting point is to prepare $N$ ions that are loaded in a trap in
some internal state with state vector $\ket0$.
One may then drive an atomic transition with natural frequency $\omega_0$
to a level $\ket1$ by applying
an appropriate Ramsey pulse with frequency $\omega$,
such that the ions are
in an equal superposition of $\ket0$ and $\ket1$.
After a free evolution for a time $t$, the probability to
find the ions in level $\ket1$ is given by
\begin{equation*}
  p = (1+ \cos((\omega-\omega_0)t) )/2.
\end{equation*}
Given such a preparation, one finds that if one estimates the
frequency $\omega_0$ with such a scheme, the uncertainty in the
estimated value is given by
\begin{equation*}
  \delta \omega_0 = (N T t)^{-1/2}.
\end{equation*}
This theoretical limit, the \emph{shot-noise} limit, can in principle be
overcome when entangling the ions initially. This idea has been
first explored in Ref.~\cite{Wineland}, where it was suggested to prepare
the ions in a $N$-particle GHZ state with state vector
  $\ket{\mathrm{GHZ}} =
  (\ket{0,\dots,0} + \ket{1,\dots,1}/\sqrt{2}$.
With such a preparation, and neglecting decoherence effects, one finds an
enhanced precision,
\begin{equation*}
  \delta \omega_0 =   (T t)^{-1/2}/N,
\end{equation*}
beating the above limit by a factor of $1/\sqrt{N}$.
Unfortunately, while the GHZ-state provides some increase in precision
in an ideal case, it is at the same time subject to decoherence processes. A more
careful analysis shows that under realistic decoherence models this
enhancement actually disappears for the GHZ state.
Notwithstanding these problems, the general idea of exploiting multi-partite
entanglement to enhance frequency-measurements can be made use
of: For example, for $N=4$ the state
\begin{align*}
  \ket\psi &= \lambda_0 \bigl( \ket{0,0,0,0} + \ket{1,1,1,1} \bigr)
  + \lambda_1 \bigl( \ket{0,0,0,1} + \ket{0,0,1,0}
  +\ket{0,1,0,0} \\
  &+  \ket{1,0,0,0} + \ket{1,1,1,0}
  + \ket{1,1,0,1} + \ket{1,0,1,1} + \ket{0,1,1,1} \bigr)\\
  &+ \lambda_2 \bigl(\ket{0,0,1,1} + \ket{0,1,0,1}+ \ket{1,0,0,1}
  + \ket{1,1,0,0} + \ket{1,0,1,0} + \ket{0,1,1,0} \bigr)
\end{align*}
can lead to an improvement of more than $6\%$, when the
probability distribution $\lambda_0,\lambda_1,\lambda_2$ is appropriately
chosen and appropriate measurements are performed~\cite{Frequency}.
For four ions, exciting experiments have been performed in the
meantime~\cite{NewWineland}, and entanglement has been used for precision spectroscopy~\cite{NewWineland2}, demonstrating that the shot noise limit can indeed be beaten with the proper use of entanglement.

{\it Acknowledgements.}---
We thank
R.~L.~Franco,
O.~G\"uhne, B.\ Kraus, and U.\ Marzolino for valuable feedback.
This work has been supported by the EPSRC, the
EU (IST-2002-38877, AQuS), the DFG (CRC QIV, project B01 of CRC 183, EI 519/9-1, EI 519/7-1),
the Templeton Foundation, Eurohorcs (EURYI), the European Research Councils (ERC CoG TAQ),
the Excellence Initiative of the German Federal and State Governments (Grant ZUK 81),
Universities Australia and DAAD's Joint Research Co-operation Scheme,
the Simons Foundation, and the AFOSR (FA9550-16-1-0082).

\bibliographystyle{plain}
{\small\bibliography{bibliography}}

\begin{thebibliography}{100}

\bibitem{AcinThreeSchmidt1}
A.~Acin, A.~Andrianov, E.~Jan{\'e}, and R.~Tarrach.
\newblock Three-qubit pure-state canonical forms.
\newblock {\em J. Phys. A}, 34:6725, 2001.

\bibitem{AcinThreeQubits}
A.~Acin, D.~Bru{\ss}, M.~Lewenstein, and A.~Sanpera.
\newblock Classification of mixed three-qubit states.
\newblock {\em Phys.\ Rev.\ Lett.}, 87:040401, 2001.

\bibitem{Acin}
A.~Acin, G.~Vidal, and J.~I. Cirac.
\newblock On the structure of a reversible entanglement generating set for
  tripartite states.
\newblock {\em Quant. Inf. Comp.}, 3:55--63, 2003.

\bibitem{PhysRevA.73.032345}
G.~Adesso, A.~Serafini, and F.~Illuminati.
\newblock Multipartite entanglement in three-mode {G}aussian states of
  continuous-variable systems: {Q}uantification, sharing structure, and
  decoherence.
\newblock {\em Phys. Rev. A}, 73:032345, 2006.

\bibitem{alicki1988symmetry}
R.~Alicki, S.~Rudnicki, and S.~Sadowski.
\newblock Symmetry properties of product states for the system of $n$ $n$-level
  atoms.
\newblock {\em J. Math. Phys.}, 29:1158--1162, 1988.

\bibitem{amico2008entanglement}
L.~Amico, R.~Fazio, A.~Osterloh, and V.~Vedral.
\newblock Entanglement in many-body systems.
\newblock {\em Rev. Mod. Phys.}, 80:517, 2008.

\bibitem{banuls_entanglement_2007}
M.-C. Ba\~{n}uls, J.~I. Cirac, and M.~M. Wolf.
\newblock Entanglement in fermionic systems.
\newblock {\em Phys. Rev. A}, 76:022311, 2007.

\bibitem{bell1964einstein}
J.~S. Bell.
\newblock {On the Einstein Podolsky Rosen paradox}.
\newblock {\em Physics}, 1:195--200, 1964.

\bibitem{BennettBiPartiteSeparable}
C.~H. Bennett, D.~P. DiVincenzo, T.~Mor, P.~W. Shor, J.~A. Smolin, and B.~M.
  Terhal.
\newblock Unextendible product bases and bound entanglement.
\newblock {\em Phys. Rev. Lett.}, 82:5385, 1999.

\bibitem{Reversible}
C.~H. Bennett, D.~P. DiVincenzo, J.~A. Smolin, and W.~K. Wootters.
\newblock Mixed-state entanglement and quantum error correction.
\newblock {\em Phys. Rev. A}, 54:3824, 1996.

\bibitem{BennettMulti}
C.~H. Bennett, S.~Popescu, D.~Rohrlich, J.~A. Smolin, and A.~V. Thapliyal.
\newblock Exact and asymptotic measures of multipartite pure-state
  entanglement.
\newblock {\em Phys. Rev. A}, 63:012307, 2000.

\bibitem{Wineland}
J.~J. Bollinger, W.~M. Itano, D.~J. Wineland, and D.~J. Heinzen.
\newblock Optimal frequency measurements with maximally correlated states.
\newblock {\em Phys. Rev. A}, 54:R4649--R4652, 1996.

\bibitem{WeinDetect}
M.~Bourennane, M.~Eibl, C.~Kurtsiefer, S~Gaertner, H.~Weinfurter, O.~G{\"u}hne,
  P.~Hyllus, D.~Bru{\ss}, M.~Lewenstein, and A.~Sanpera.
\newblock Experimental detection of multipartite entanglement using witness
  operators.
\newblock {\em Phys. Rev. Lett.}, 92:087902, 2004.

\bibitem{Zeil}
D.~Bouwmeester, J.-W. Pan, D.~I. Matthew, H.~Weinfurter, and A.~Zeilinger.
\newblock Observation of three-photon {G}reenberger-{H}orne-{Z}eilinger
  entanglement.
\newblock {\em Phys. Rev. Lett.}, 82:1345, 1999.

\bibitem{brandao2016mathematics}
F.~G. S.~L. Brandao, M.~Christandl, A.~W. Harrow, and M.~Walter.
\newblock {T}he {M}athematics of {E}ntanglement.
\newblock 2016.
\newblock Lecture notes.

\bibitem{Fer}
F.~G. S.~L. Brandao and R.~O. Vianna.
\newblock Separable multipartite mixed states: operational asymptotically
  necessary and sufficient conditions.
\newblock {\em Phys. Rev. Lett.}, 93:220503, 2004.

\bibitem{bravyi2006ghz}
S.~Bravyi, D.~Fattal, and D.~Gottesman.
\newblock {GHZ} extraction yield for multipartite stabilizer states.
\newblock {\em J. Math. Phys.}, 47:062106, 2006.

\bibitem{brennen2008measurement}
G.~K. Brennen and A.~Miyake.
\newblock Measurement-based quantum computer in the gapped ground state of a
  two-body {H}amiltonian.
\newblock {\em Phys. Rev. Lett.}, 101:010502, 2008.

\bibitem{briand2003complete}
E.~Briand, J.-G. Luque, and J.-Y. Thibon.
\newblock A complete set of covariants of the four qubit system.
\newblock {\em J. Phys. A}, 36:9915, 2003.

\bibitem{broecker2015entanglement}
P.~Broecker and S.~Trebst.
\newblock Entanglement and the fermion sign problem in auxiliary field quantum
  monte carlo simulations.
\newblock {\em 1511.02878}, 2015.

\bibitem{buhrman2016nondeterministic}
H.~Buhrman, M.~Christandl, and J.~Zuiddam.
\newblock Nondeterministic quantum communication complexity: the cyclic
  equality game and iterated matrix multiplication.
\newblock 2016.
\newblock arXiv:1603.03757.

\bibitem{cadney2012infinitely}
J.~Cadney, N.~Linden, and A.~Winter.
\newblock Infinitely many constrained inequalities for the von {N}eumann
  entropy.
\newblock {\em IEEE Trans. Inf. Th.}, 58:3657--3663, 2012.

\bibitem{Metrology}
P.~Cappellaro, J.~Emerson, N.~Boulant, C.~Ramanathan, S.~Lloyd, and D.~G Cory.
\newblock Entanglement assisted metrology.
\newblock {\em Phys. Rev. Lett.}, 94:020502, 2005.

\bibitem{CarteretMultiSchmidt}
H.~A. Carteret, A.~Higuchi, and A.~Sudbery.
\newblock Multipartite generalization of the {S}chmidt decomposition.
\newblock {\em J.\ Math.\ Phys}, 41:7932--7939, 2000.

\bibitem{chitambar2008tripartite}
E.~Chitambar, R.~Duan, and Y.~Shi.
\newblock Tripartite entanglement transformations and tensor rank.
\newblock {\em Phys. Rev. Lett.}, 101:140502, 2008.

\bibitem{chitambar2014everything}
E.~Chitambar, D.~Leung, L.~Man{\v{c}}inska, M.~Ozols, and A.~Winter.
\newblock Everything you always wanted to know about {LOCC} (but were afraid to
  ask).
\newblock {\em Commun. Math. Phys.}, 328:303--326, 2014.

\bibitem{christandl2006spectra}
M.~Christandl and G.~Mitchison.
\newblock The spectra of quantum states and the {K}ronecker coefficients of the
  symmetric group.
\newblock {\em Commun. Math. Phys.}, 261:789--797, 2006.

\bibitem{christandl2012recoupling}
M.~Christandl, M.~B. {\c{S}}ahino{\u{g}}lu, and M.~Walter.
\newblock Recoupling coefficients and quantum entropies.
\newblock 2012.

\bibitem{clauser1969proposed}
J.~F. Clauser, M.~A. Horne, A.~Shimony, and R.~A. Holt.
\newblock Proposed experiment to test local hidden-variable theories.
\newblock {\em Phys. Rev. Lett.}, 23:880, 1969.

\bibitem{PhysRevLett.83.648}
R.~Cleve, D.~Gottesman, and H.-K. Lo.
\newblock How to share a quantum secret.
\newblock {\em Phys. Rev. Lett.}, 83:648--651, 1999.

\bibitem{Tangle}
V.~Coffman, J.~Kundu, and W.~K. Wootters.
\newblock Distributed entanglement.
\newblock {\em Phys. Rev. A}, 61:052306, 2000.

\bibitem{PhysRevA.73.012309}
M.~Cramer, J.~Eisert, M.~B. Plenio, and J.~Dreissig.
\newblock Entanglement-area law for general bosonic harmonic lattice systems.
\newblock {\em Phys. Rev. A}, 73:012309, 2006.

\bibitem{SecureFunctionEvaluation}
C.~Creapeau, D.~Gottesman, and A.~Smith.
\newblock Secure multi-party quantum computing.
\newblock {\em Proc. 34th ACM Stoc.}, 643, 2002.

\bibitem{Darmawan}
A.~S. Darmawan, G.~K. Brennen, and S.~D. Bartlett.
\newblock Measurement-based quantum computation in a {2D} phase of matter.
\newblock {\em New J. Phys.}, 14:013023, 2012.

\bibitem{KrausMaximallyEntangled}
J.~I. de~Vicente, C.~Spee, and B.~Kraus.
\newblock The maximally entangled set of multipartite quantum states.
\newblock {\em Phys. Rev. Lett.}, 111:110502, 2013.

\bibitem{ding2016conditional}
D.~Ding, P.~Hayden, and M.~Walter.
\newblock Conditional mutual information of bipartite unitaries and scrambling.
\newblock {\em J. High Energy Phys.}, 12:145, 2016.

\bibitem{Doh}
A.~C. Doherty, P.~A. Parrilo, and F.~M. Spedalieri.
\newblock Distinguishing separable and entangled states.
\newblock {\em Phys. Rev. Lett.}, 88:187904, 2002.

\bibitem{Uniqueness}
M.~J. Donald, M.~Horodecki, and O.~Rudolph.
\newblock The uniqueness theorem for entanglement measures.
\newblock {\em J. Math. Phys.}, 43:4252, 2002.

\bibitem{dougherty2006six}
R.~Dougherty, C.~Freiling, and K.~Zeger.
\newblock Six new non-{S}hannon information inequalities.
\newblock In {\em Proc. 2006 ISIT}, pages 233--236. IEEE, 2006.

\bibitem{PhysRevA.73.052323}
M.~R. Dowling, A.~C. Doherty, and H.~M. Wiseman.
\newblock Entanglement of indistinguishable particles in condensed-matter
  physics.
\newblock {\em Phys. Rev. A}, 73:052323, 2006.

\bibitem{DuerMultiQubitClassification}
W.~D{\"u}r and J.~I. Cirac.
\newblock {Classification of multiqubit mixed states: Separability and
  distillability properties}.
\newblock {\em Phys. Rev. A}, 61:042314, 2000.

\bibitem{DuerMixedClassification}
W.~D{\"u}r, J.~I. Cirac, and R.~Tarrach.
\newblock Separability and distillability of multiparticle quantum systems.
\newblock {\em Phys. Rev. Lett.}, 83:3562, 1999.

\bibitem{DuerThreeQubits}
W.~D{\"u}r, G.~Vidal, and J.~I. Cirac.
\newblock Three qubits can be entangled in two inequivalent ways.
\newblock {\em Phys. Rev. A}, 62:062314, 2000.

\bibitem{eckert_quantum_2002}
K.~Eckert, J.~Schliemann, D.~Bru\ss{}, and M.~Lewenstein.
\newblock Quantum {Correlations} in {Systems} of {Indistinguishable}
  {Particles}.
\newblock {\em Ann. Phys.}, 299:88--127, 2002.

\bibitem{Wein}
M.~Eibl, N.~Kiesel, M.~Bourennane, C.~Kurtsiefer, and H.~Weinfurter.
\newblock Experimental realization of a three-qubit entangled {$W$} state.
\newblock {\em Phys. Rev. Lett.}, 92:077901, 2004.

\bibitem{einstein1935can}
A.~Einstein, B.~Podolsky, and N.~Rosen.
\newblock Can quantum-mechanical description of physical reality be considered
  complete?
\newblock {\em Phys. Rev.}, 47:777, 1935.

\bibitem{PhD}
J.~Eisert.
\newblock {\em Entanglement in quantum information theory}.
\newblock PhD thesis, University of Potsdam, 2001.

\bibitem{Schmidt}
J.~Eisert and H.-J. Briegel.
\newblock Schmidt measure as a tool for quantifying multiparticle entanglement.
\newblock {\em Phys. Rev. A}, 64:022306, 2001.

\bibitem{AreaReview}
J.~Eisert, M.~Cramer, and M.~B. Plenio.
\newblock Area laws for the entanglement entropy.
\newblock {\em Rev. Mod. Phys.}, 82:277, 2010.

\bibitem{Hierarchy}
J.~Eisert, P.~Hyllus, O.~G{\"u}hne, and M.~Curty.
\newblock Complete hierarchies of efficient approximations to problems in
  entanglement theory.
\newblock {\em Phys. Rev. A}, 70:062317, 2004.

\bibitem{GaussianIntro}
J.~Eisert and M.~B. Plenio.
\newblock Introduction to the basics of entanglement theory in
  continuous-variable systems.
\newblock {\em Int. J. Quant. Inf.}, 1:479, 2003.

\bibitem{GaussianQuantumMarginals}
J.~Eisert, T.~Tyc, T.~Rudolph, and B.~C. Sanders.
\newblock Gaussian quantum marginal problem.
\newblock {\em Commun. Math. Phys.}, 280:263, 2008.

\bibitem{eltschka2012multipartite}
C.~Eltschka, T.~Bastin, A.~Osterloh, and J.~Siewert.
\newblock Multipartite-entanglement monotones and polynomial invariants.
\newblock {\em Phys. Rev. A}, 85:022301, 2012.

\bibitem{Ugo}
P.~Facchi, G.~Florio, U.~Marzolino, G.~Parisi, and S.~Pascazio.
\newblock Multipartite entanglement and frustration.
\newblock {\em New J. Phys.}, 12, 2010.

\bibitem{PhysRevA.77.060304}
P.~Facchi, G.~Florio, G.~Parisi, and S.~Pascazio.
\newblock Maximally multipartite entangled states.
\newblock {\em Phys. Rev. A}, 77:060304, 2008.

\bibitem{PhysRevA.74.042331}
P.~Facchi, G.~Florio, and S.~Pascazio.
\newblock Probability-density-function characterization of multipartite
  entanglement.
\newblock {\em Phys. Rev. A}, 74:042331, 2006.

\bibitem{raey}
M.~Fannes, B.~Nachtergaele, and R.~F. Werner.
\newblock Finitely correlated states on quantum spin chains.
\newblock {\em Commun. Math. Phys.}, 144:443--490, 1992.

\bibitem{fawzi2015quantum}
O.~Fawzi and R.~Renner.
\newblock {Quantum conditional mutual information and approximate Markov
  chains}.
\newblock {\em Commun. Math. Phys.}, 340:575--611, 2015.

\bibitem{franco_quantum_2016}
R.~L. Franco and G.~Compagno.
\newblock Quantum entanglement of identical particles by standard
  information-theoretic notions.
\newblock {\em Sci. Rep.}, 6:20603, 2016.

\bibitem{VirmaniMREGS}
E.~F. Galvao, M.~B. Plenio, and S~Virmani.
\newblock Tripartite entanglement and quantum relative entropy.
\newblock {\em J. Phys. A}, 33:8809, 2000.

\bibitem{gharibian2010strong}
S.~Gharibian.
\newblock Strong {NP}-hardness of the quantum separability problem.
\newblock {\em Quant. Inf. Comp.}, 10:343--360, 2010.

\bibitem{PhysRevLett.115.020502}
M.~Gimeno-Segovia, P.~Shadbolt, D.~E. Browne, and T.~Rudolph.
\newblock {From three-photon Greenberger-Horne-Zeilinger states to ballistic
  universal quantum computation}.
\newblock {\em Phys. Rev. Lett.}, 115:020502, 2015.

\bibitem{PhysRevA.68.062108}
P.~Giorda and P.~Zanardi.
\newblock Mode entanglement and entangling power in bosonic graphs.
\newblock {\em Phys. Rev. A}, 68:062108, 2003.

\bibitem{Metrology2}
V.~Giovannetti, S.~Lloyd, and L.~Maccone.
\newblock Quantum-enhanced measurements: beating the standard quantum limit.
\newblock {\em Science}, 306:1330--1336, 2004.

\bibitem{gottesman96stabilizer}
D.~Gottesman.
\newblock {\em Stabilizer codes and quantum error correction}.
\newblock PhD thesis, Caltech, 1996.

\bibitem{GourAME}
G.~Gour and N.~R. Wallach.
\newblock All maximally entangled four-qubit states.
\newblock {\em J. Math. Phys.}, 51:112201, 2010.

\bibitem{GourWallach}
G.~Gour and N.~R. Wallach.
\newblock Classification of multipartite entanglement of all finite
  dimensionality.
\newblock {\em Phys. Rev. Lett.}, 111:060502, 2013.

\bibitem{GrasslInvariants}
M.~Grassl, M.~R{\"o}tteler, and T.~Beth.
\newblock Computing local invariants of quantum-bit systems.
\newblock {\em Phys.\ Rev.\ A}, 58:1833, 1998.

\bibitem{greenberger1989going}
D.~M. Greenberger, M.~A. Horne, and A.~Zeilinger.
\newblock Going beyond {Bell}'s theorem.
\newblock In {\em Bell's theorem, quantum theory and conceptions of the
  universe}, pages 69--72. Springer, 1989.

\bibitem{gross2007novel}
D.~Gross and J.~Eisert.
\newblock Novel schemes for measurement-based quantum computation.
\newblock {\em Phys. Rev. Lett.}, 98:220503, 2007.

\bibitem{gross2007measurement}
D.~Gross, J.~Eisert, N.~Schuch, and D.~Perez-Garcia.
\newblock Measurement-based quantum computation beyond the one-way model.
\newblock {\em Phys. Rev. A}, 76:052315, 2007.

\bibitem{gross2009most}
D.~Gross, S.~T. Flammia, and J.~Eisert.
\newblock Most quantum states are too entangled to be useful as computational
  resources.
\newblock {\em Phys. Rev. Lett.}, 102:190501, 2009.

\bibitem{gross2013stabilizer}
D.~Gross and M.~Walter.
\newblock Stabilizer information inequalities from phase space distributions.
\newblock {\em J. Math. Phys.}, 54:082201, 2013.

\bibitem{Guehne}
O.~G{\"u}hne, P.~Hyllus, D.~Bru{\ss}, A.~Ekert, M.~Lewenstein, C.~Macchiavello,
  and A.~Sanpera.
\newblock Detection of entanglement with few local measurements.
\newblock {\em Phys. Rev. A}, 66:062305, 2002.

\bibitem{guhne2009entanglement}
O.~G{\"u}hne and G.~T{\'o}th.
\newblock Entanglement detection.
\newblock {\em Phys. Rep.}, 474:1--75, 2009.

\bibitem{guhne2005bell}
O.~G{\"u}hne, G.~T{\'o}th, P.~Hyllus, and H.~J. Briegel.
\newblock Bell inequalities for graph states.
\newblock {\em Phys. Rev. Lett.}, 95:120405, 2005.

\bibitem{Gurvits}
L.~Gurvits.
\newblock {Classical deterministic complexity of Edmonds' Problem and quantum
  entanglement}.
\newblock In {\em Proc. STOC}, pages 10--19. ACM., 2003.

\bibitem{OneD}
M.~B. Hastings.
\newblock An area law for one dimensional quantum systems.
\newblock {\em J. Stat. Mech.}, page P08024, 2007.

\bibitem{hayashi2002quantum}
M.~Hayashi and K.~Matsumoto.
\newblock Quantum universal variable-length source coding.
\newblock {\em Phys. Rev. A}, 66:022311, 2002.

\bibitem{hayden2016holographic}
P.~Hayden, Sepehr Nezami, X.-L. Qi, N.~Thomas, M.~Walter, and Z.~Yang.
\newblock Holographic duality from random tensor networks.
\newblock {\em J. High Energy Phys.}, 11:9, 2016.

\bibitem{Graphs}
M.~Hein, J.~Eisert, and H.~J. Briegel.
\newblock Multiparty entanglement in graph states.
\newblock {\em Phys. Rev. A}, 69:062311, 2004.

\bibitem{PhysRevA.86.052335}
W.~Helwig, W.~Cui, J.~I. Latorre, A.~Riera, and H.-K. Lo.
\newblock Absolute maximal entanglement and quantum secret sharing.
\newblock {\em Phys. Rev. A}, 86:052335, 2012.

\bibitem{hill1997entanglement}
I.~Hill and W.~K. Wootters.
\newblock Entanglement of a pair of quantum bits.
\newblock {\em Phys. Rev. Lett.}, 78:5022, 1997.

\bibitem{PhysRevA.59.1829}
M.~Hillery, V.~Bu\ifmmode~\check{z}\else \v{z}\fi{}ek, and A.~Berthiaume.
\newblock Quantum secret sharing.
\newblock {\em Phys. Rev. A}, 59:1829--1834, 1999.

\bibitem{holevo1973bounds}
A.~S. Holevo.
\newblock Bounds for the quantity of information transmitted by a quantum
  communication channel.
\newblock {\em Prob. Per. Inf.}, 9:3--11, 1973.

\bibitem{HorodeckiEntanglement}
R.~Horodecki, P.~Horodecki, M.~Horodecki, and K.~Horodecki.
\newblock Quantum entanglement.
\newblock {\em Rev. Mod. Phys.}, 81:865, 2009.

\bibitem{GuehneAME}
F.~Huber, O.~G{\"u}hne, and J.~Siewert.
\newblock Absolutely maximally entangled states of seven qubits do not exist.
\newblock arXiv:1608.06228.

\bibitem{Frequency}
S.~F. Huelga, C.~Macchiavello, T.~Pellizzari, A.~K. Ekert, M.~B. Plenio, and
  J.~I. Cirac.
\newblock Improvement of frequency standards with quantum entanglement.
\newblock {\em Phys. Rev. Lett.}, 79:3865, 1997.

\bibitem{Bruss}
F.~Hulpke and D.~Bru{\ss}.
\newblock A two-way algorithm for the entanglement problem.
\newblock {\em J. Phys. A}, 38:5573, 2005.

\bibitem{PhysRevA.86.012337}
P.~Hyllus, L.~Pezz\'e, A.~Smerzi, and G.~T\'oth.
\newblock Entanglement and extreme spin squeezing for a fluctuating number of
  indistinguishable particles.
\newblock {\em Phys. Rev. A}, 86:012337, 2012.

\bibitem{keyl2001estimating}
M.~Keyl and R.~F. Werner.
\newblock Estimating the spectrum of a density operator.
\newblock {\em Phys. Rev. A}, 64:052311, 2001.

\bibitem{QuantumInternet}
H.~J. Kimble.
\newblock The quantum internet.
\newblock {\em Nature}, 453:1023--1030, 2008.

\bibitem{kitaev2006topological}
A.~Kitaev and J.~Preskill.
\newblock Topological entanglement entropy.
\newblock {\em Phys. Rev. Lett.}, 96:110404, 2006.

\bibitem{kitaev2003fault}
A.~Y. Kitaev.
\newblock Fault-tolerant quantum computation by anyons.
\newblock {\em Ann. Phys.}, 303:2--30, 2003.

\bibitem{klyachko2007dynamical}
A.~Klyachko.
\newblock Dynamical symmetry approach to entanglement.
\newblock {\em Proc. NATO Adv. Study Inst. Phys. Theo. Comp. Sci.}, 7:25, 2007.

\bibitem{PhysRevA.82.032121}
B.~Kraus.
\newblock Local unitary equivalence and entanglement of multipartite pure
  states.
\newblock {\em Phys. Rev. A}, 82:032121, 2010.

\bibitem{PhysRevLett.104.020504}
B.~Kraus.
\newblock Local unitary equivalence of multipartite pure states.
\newblock {\em Phys. Rev. Lett.}, 104:020504, 2010.

\bibitem{landsberg2012tensors}
J.~M. Landsberg.
\newblock {\em Tensors: geometry and applications}, volume 128.
\newblock American Mathematical Society, 2012.

\bibitem{laskowski2012incompatible}
W.~Laskowski, M.~Markiewicz, T.~Paterek, and M.~Wie{\'s}niak.
\newblock Incompatible local hidden-variable models of quantum correlations.
\newblock {\em Phys. Rev. A}, 86:032105, 2012.

\bibitem{leifer2004measuring}
M.~S. Leifer, N.~Linden, and A.~Winter.
\newblock Measuring polynomial invariants of multiparty quantum states.
\newblock {\em Phys. Rev. A}, 69:052304, 2004.

\bibitem{levin2006detecting}
M.~Levin and X.-G. Wen.
\newblock Detecting topological order in a ground state wave function.
\newblock {\em Phys. Rev. Lett.}, 96:110405, 2006.

\bibitem{lieb1973fundamental}
E.~H. Lieb and M.~B. Ruskai.
\newblock A fundamental property of quantum-mechanical entropy.
\newblock {\em Phys. Rev. Lett.}, 30:434, 1973.

\bibitem{lekheng}
L.-H. Lim and C.~Hillar.
\newblock Most tensor problems are {NP}-hard.
\newblock {\em J. ACM.}, 60, 2013.

\bibitem{linden2013quantum}
N.~Linden, F.~Mat{\'u}{\v{s}}, M.~B. Ruskai, and A.~Winter.
\newblock The quantum entropy cone of stabiliser states.
\newblock In {\em Proc. 8th TQC Guelph}, volume~22 of {\em LIPICS}, pages
  270--284, 2013.

\bibitem{LindenMultiEntanglement}
N.~Linden and S.~Popescu.
\newblock On multi-particle entanglement.
\newblock {\em Fortschr. Phys.}, 46:567--578, 1998.

\bibitem{LindenMREGS}
N.~Linden, S.~Popescu, B.~Schumacher, and M.~Westmoreland.
\newblock Reversibility of local transformations of multiparticle entanglement.
\newblock 1999.
\newblock arXiv:quant-ph/9912039.

\bibitem{LindenNonLocalDensities}
N.~Linden, S.~Popescu, and A.~Sudbery.
\newblock Nonlocal parameters for multiparticle density matrices.
\newblock {\em Phys. Rev. Lett.}, 83:243, 1999.

\bibitem{linden2005new}
N.~Linden and A.~Winter.
\newblock A new inequality for the von {N}eumann entropy.
\newblock {\em Commun. Math. Phys.}, 259:129--138, 2005.

\bibitem{luque2006algebraic}
J.-G. Luque and J.-Y. Thibon.
\newblock Algebraic invariants of five qubits.
\newblock {\em J. Phys. A}, 39:371--377, 2006.

\bibitem{luque2007unitary}
J.-G. Luque, J.-Y. Thibon, and F.~Toumazet.
\newblock Unitary invariants of qubit systems.
\newblock {\em Math. Struct. Comp. Sci.}, 17:1133--1151, 2007.

\bibitem{matus2007infinitely}
F.~Matus.
\newblock Infinitely many information inequalities.
\newblock In {\em Proc. 2007 ISIT}, pages 41--44. IEEE, 2007.

\bibitem{mermin1990extreme}
N.~D. Mermin.
\newblock Extreme quantum entanglement in a superposition of macroscopically
  distinct states.
\newblock {\em Phys. Rev. Lett.}, 65:1838, 1990.

\bibitem{NewWineland2}
V.~Meyer, M.~A. Rowe, D.~Kielpinski, C.~A. Sackett, W.~M. Itano, C.~Monroe, and
  D.~J. Wineland.
\newblock Experimental demonstration of entanglement-enhanced rotation angle
  estimation using trapped ions.
\newblock {\em Phys. Rev. Lett.}, 86:5870, 2001.

\bibitem{miyake_classification_2003}
A.~Miyake.
\newblock Classification of multipartite entangled states by multidimensional
  determinants.
\newblock {\em Phys. Rev. A}, 67:012108, 2003.

\bibitem{miyake2011quantum}
A.~Miyake.
\newblock Quantum computational capability of a 2d valence bond solid phase.
\newblock {\em Ann. Phys.}, 326:1656--1671, 2011.

\bibitem{SecuritySecretSharing}
A.~C.~A. Nascimento, P.~Tuyls, A.~Winter, H.~Imai, and J.~M{\"u}ller-Quade.
\newblock A quantum information theoretical model for quantum secret sharing
  schemes.
\newblock 2003.
\newblock arXiv:quant-ph/0311136.

\bibitem{nezami2016multipartite}
S.~Nezami and M.~Walter.
\newblock {Multipartite entanglement in stabilizer tensor networks}.
\newblock 2016.
\newblock 1608.02595.

\bibitem{nielsen1999conditions}
M.~A. Nielsen.
\newblock Conditions for a class of entanglement transformations.
\newblock {\em Phys. Rev. Lett.}, 83:436--439, 1999.

\bibitem{nielsen2000quantum}
M.~A. Nielsen and I.~L. Chuang.
\newblock {\em Quantum information and quantum computation}.
\newblock Cambridge University Press, 2000.

\bibitem{Orus-AnnPhys-2014}
R.~Or\'us.
\newblock A practical introduction to tensor networks: Matrix product states
  and projected entangled pair states.
\newblock {\em Ann. Phys.}, 349:117--158, 2014.

\bibitem{orus2014topological}
R.~Or{\'u}s, T.-C. Wei, O.~Buerschaper, and A.~Garc{\'\i}a-Saez.
\newblock Topological transitions from multipartite entanglement with tensor
  networks: a procedure for sharper and faster characterization.
\newblock {\em Phys. Rev. Lett.}, 113:257202, 2014.

\bibitem{osterloh2005constructing}
A.~Osterloh and J.~Siewert.
\newblock Constructing {$N$}-qubit entanglement monotones from antilinear
  operators.
\newblock {\em Phys. Rev. A}, 72:012337, 2005.

\bibitem{osterloh2006entanglement}
A.~Osterloh and J.~Siewert.
\newblock Entanglement monotones and maximally entangled states in multipartite
  qubit systems.
\newblock {\em Int. J. Quant. Inf.}, 4:531--540, 2006.

\bibitem{palazuelos2016survey}
C.~Palazuelos and T.~Vidick.
\newblock Survey on nonlocal games and operator space theory.
\newblock {\em J. Math. Phys.}, 57:015220, 2016.

\bibitem{pan2000experimental}
J.-W. Pan, D.~Bouwmeester, D.~L. Matthew, H.~Weinfurter, and A.~Zeilinger.
\newblock Experimental test of quantum nonlocality in three-photon
  {G}reenberger-{H}orne-{Z}eilinger entanglement.
\newblock {\em Nature}, 403:515--519, 2000.

\bibitem{pastawski2015holographic}
F.~Pastawski, B.~Yoshida, D.~Harlow, and J.~Preskill.
\newblock Holographic quantum error-correcting codes: {T}oy models for the
  bulk/boundary correspondence.
\newblock {\em J. High Energy Phys.}, 06:149, 2015.

\bibitem{pippenger2003inequalities}
N.~Pippenger.
\newblock The inequalities of quantum information theory.
\newblock {\em IEEE Trans. Inf. Th.}, 49:773--789, 2003.

\bibitem{NotConvex}
M.~B. Plenio.
\newblock Logarithmic negativity: A full entanglement monotone that is not
  convex.
\newblock {\em Phys. Rev. Lett.}, 95:090503, 2005.

\bibitem{RelEntMulti}
M.~B. Plenio and V.~Vedral.
\newblock Bounds on relative entropy of entanglement for multi-party systems.
\newblock {\em J. Phys. A}, 34:6997, 2001.

\bibitem{plenio2007introduction}
M.~B. Plenio and S.~Virmani.
\newblock An introduction to entanglement measures.
\newblock {\em Quant. Inf. Comp.}, 7:1--51, 2007.

\bibitem{PhysRevA.67.012322}
M.~Plesch and V.~Bu\ifmmode~\check{z}\else \v{z}\fi{}ek.
\newblock Entangled graphs: Bipartite entanglement in multiqubit systems.
\newblock {\em Phys. Rev. A}, 67:012322, 2003.

\bibitem{PhysRevB.86.125441}
F.~Pollmann and A.~M. Turner.
\newblock Detection of symmetry-protected topological phases in one dimension.
\newblock {\em Phys. Rev. B}, 86:125441, 2012.

\bibitem{Localizable}
M.~Popp, F.~Verstraete, M.~A. Mart{\'\i}n-Delgado, and J.~I. Cirac.
\newblock Localizable entanglement.
\newblock {\em Phys. Rev. A}, 71:042306, 2005.

\bibitem{qi2013exact}
X.-L. Qi.
\newblock Exact holographic mapping and emergent space-time geometry.
\newblock 2013.
\newblock arXiv:1309.6282.

\bibitem{RainsInvariants}
E.~M. Rains.
\newblock Nonbinary quantum codes.
\newblock {\em IEEE Trans. Inf. Th.}, 45:1827--1832, 1999.

\bibitem{RainsCodes}
E.~M. Rains.
\newblock Quantum codes of minimum distance two.
\newblock {\em IEEE Trans. Inf. Th.}, 45:266--271, 1999.

\bibitem{Oneway}
R.~Raussendorf and H.~J. Briegel.
\newblock A one-way quantum computer.
\newblock {\em Phys. Rev. Lett.}, 86:5188, 2001.

\bibitem{Blatt}
C.~F. Roos, M.~Riebe, H.~H{\"a}ffner, W.~H{\"a}nsel, J.~Benhelm, G.~P.~T.
  Lancaster, C.~Becher, F.~Schmidt-Kaler, and R.~Blatt.
\newblock Control and measurement of three-qubit entangled states.
\newblock {\em Science}, 304:1478--1480, 2004.

\bibitem{NewWineland}
C.~A. Sackett, D.~Kielpinski, B.~E. King, C.~Langer, V~Meyer, C.~J. Myatt,
  M.~Rowe, Q.~A. Turchette, W.~M. Itano, D.~J. Wineland, et~al.
\newblock Experimental entanglement of four particles.
\newblock {\em Nature}, 404:56--259, 2000.

\bibitem{sawicki2012critical}
A.~Sawicki, M.~Oszmaniec, and M.~Ku{\'s}.
\newblock Critical sets of the total variance can detect all stochastic local
  operations and classical communication classes of multiparticle entanglement.
\newblock {\em Phys. Rev. A}, 86:040304, 2012.

\bibitem{sawicki2014convexity}
A.~Sawicki, M.~Oszmaniec, and M.~Ku{\'s}.
\newblock Convexity of momentum map, {M}orse index, and quantum entanglement.
\newblock {\em Rev. Math. Phys.}, 26:1450004, 2014.

\bibitem{sawicki2013pure}
A.~Sawicki, M.~Walter, and M.~Ku{\'s}.
\newblock When is a pure state of three qubits determined by its
  single-particle reduced density matrices?
\newblock {\em J. Phys. A}, 46:055304, 2013.

\bibitem{Schlinge}
D.~Schlingemann.
\newblock Logical network implementation for cluster states and graph codes.
\newblock {\em Quant. Inf. Comp.}, pages 431--449, 2003.

\bibitem{MPSRev}
U.~Schollw\"ock.
\newblock The density-matrix renormalization group in the age of matrix product
  states.
\newblock {\em Ann. Phys.}, 326:96, 2011.

\bibitem{1010.3732}
N.~Schuch, D.~Perez-Garcia, and J.~I. Cirac.
\newblock Classifying quantum phases using matrix product states and projected
  entangled pair states.
\newblock {\em Phys. Rev. B}, 84:165139, 2011.

\bibitem{schuch_quantum_2004}
N.~Schuch, F.~Verstraete, and J.~I. Cirac.
\newblock Quantum entanglement theory in the presence of superselection rules.
\newblock {\em Phys. Rev. A}, 70:042310, 2004.

\bibitem{SchuchApprox}
N.~Schuch, M.~M. Wolf, F.~Verstraete, and J.~I. Cirac.
\newblock {Entropy scaling and simulability by Matrix Product States}.
\newblock {\em Phys. Rev. Lett.}, 100:030504, 2008.

\bibitem{smith2006typical}
G.~Smith and D.~Leung.
\newblock Typical entanglement of stabilizer states.
\newblock {\em Phys. Rev. A}, 74:062314, 2006.

\bibitem{Spee}
C.~Spee, J.~I. de~Vicente, and B.~Kraus.
\newblock The maximally entangled set of 4-qubit states.
\newblock {\em J. Math. Phys.}, 57:052201, 2016.

\bibitem{swingle2012constructing}
B.~Swingle.
\newblock Constructing holographic spacetimes using entanglement
  renormalization.
\newblock 2012.
\newblock arXiv:1209.3304.

\bibitem{swingle2012entanglement}
B.~Swingle.
\newblock Entanglement renormalization and holography.
\newblock {\em Phys. Rev. D}, 86:065007, 2012.

\bibitem{toth2007optimal}
G.~T{\'o}th, C.~Knapp, O.~G{\"u}hne, and H.~J. Briegel.
\newblock Optimal spin squeezing inequalities detect bound entanglement in spin
  models.
\newblock {\em Phys. Rev. Lett.}, 99:250405, 2007.

\bibitem{van2006universal}
M.~Van~den Nest, A.~Miyake, W.~D{\"u}r, and H.~J Briegel.
\newblock Universal resources for measurement-based quantum computation.
\newblock {\em Phys. Rev. Lett.}, 97:150504, 2006.

\bibitem{verstraete_quantum_2003}
F.~Verstraete and J.~I. Cirac.
\newblock Quantum {Nonlocality} in the {Presence} of {Superselection} {Rules}
  and {Data} {Hiding} {Protocols}.
\newblock {\em Phys. Rev. Lett.}, 91:010404, 2003.

\bibitem{verstraete_normal_2003}
F.~Verstraete, J.~Dehaene, and B.~De~Moor.
\newblock Normal forms and entanglement measures for multipartite quantum
  states.
\newblock {\em Phys. Rev. A}, 68:012103, 2003.

\bibitem{Frank}
F.~Verstraete, J.~Dehaene, B.~De~Moor, and H.~Verschelde.
\newblock Four qubits can be entangled in nine different ways.
\newblock {\em Phys. Rev. A}, 65:052112, 2002.

\bibitem{verstraete2001local}
F.~Verstraete, J.~Dehaene, and B.~DeMoor.
\newblock Local filtering operations on two qubits.
\newblock {\em Phys. Rev. A}, 64:010101, 2001.

\bibitem{Monotones}
G.~Vidal.
\newblock Entanglement monotones.
\newblock {\em J. Mod. Opt.}, 47:355, 2000.

\bibitem{vidal2007entanglement}
G.~Vidal.
\newblock Entanglement renormalization.
\newblock {\em Phys. Rev. Lett.}, 99:220405, 2007.

\bibitem{vidal2008class}
G.~Vidal.
\newblock Class of quantum many-body states that can be efficiently simulated.
\newblock {\em Phys. Rev. Lett.}, 101:110501, 2008.

\bibitem{ReversibleVidal}
G.~Vidal, W.~D{\"u}r, and J.~I. Cirac.
\newblock Reversible combination of inequivalent kinds of multipartite
  entanglement.
\newblock {\em Phys. Rev. Lett.}, 85:658, 2000.

\bibitem{vrana2015asymptotic}
P.~Vrana and M.~Christandl.
\newblock Asymptotic entanglement transformation between {W} and {GHZ} states.
\newblock {\em J. Math. Phys.}, 56:022204, 2015.

\bibitem{vrana2016entanglement}
P.~Vrana and M.~Christandl.
\newblock Entanglement distillation from {G}reenberger-{H}orne-{Z}eilinger
  shares.
\newblock 2016.

\bibitem{walter2014phd}
M.~Walter.
\newblock {\em Multipartite Quantum States and their Marginals}.
\newblock PhD thesis, ETH Zurich, 2014.

\bibitem{walter2013entanglement}
M.~Walter, B.~Doran, D.~Gross, and M.~Christandl.
\newblock Entanglement polytopes: multiparticle entanglement from
  single-particle information.
\newblock {\em Science}, 340:1205--1208, 2013.

\bibitem{webb_clifford_2015}
Z.~Webb.
\newblock The {Clifford} group forms a unitary 3-design.
\newblock 2015.
\newblock arXiv:1510.02769.

\bibitem{RevModPhys.84.621}
C.~Weedbrook, S.~Pirandola, R.~Garc\'{\i}a-Patr\'on, N.~J. Cerf, T.~C. Ralph,
  J.~H. Shapiro, and S.~Lloyd.
\newblock Gaussian quantum information.
\newblock {\em Rev. Mod. Phys.}, 84:621--669, 2012.

\bibitem{Geometric}
T.-C. Wei and P.~M. Goldbart.
\newblock Geometric measure of entanglement and applications to bipartite and
  multipartite quantum states.
\newblock {\em Phys. Rev. A}, 68:042307, 2003.

\bibitem{werner1989quantum}
R.~F. Werner.
\newblock Quantum states with {E}instein-{P}odolsky-{R}osen correlations
  admitting a hidden-variable model.
\newblock {\em Phys. Rev. A}, 40:4277, 1989.

\bibitem{DMRGWhite92}
S.~R. White.
\newblock Density matrix formulation for quantum renormalization groups.
\newblock {\em Phys. Rev. Lett.}, 69:2863, 1992.

\bibitem{wilde2011classical}
M.~M. Wilde.
\newblock {\em From classical to quantum {S}hannon theory}.
\newblock Cambridge University Press, 2011.

\bibitem{WineOld}
D.~J. Wineland, J.~J. Bollinger, W.~M. Itano, and D.~J. Heinzen.
\newblock Squeezed atomic states and projection noise in spectroscopy.
\newblock {\em Phys. Rev. A}, 50:67, 1994.

\bibitem{winter2012stronger}
A.~Winter and K.~Li.
\newblock A stronger subadditivity relation.
\newblock 2012.
\newblock
  \href{https://sites.google.com/site/derwinter/news/stronger_subadditivity.pdf}{https://sites.google.com/site/derwinter/news/stronger\_subadditivity.pdf}.

\bibitem{wootters1998entanglement}
W.~K. Wootters.
\newblock Entanglement of formation of an arbitrary state of two qubits.
\newblock {\em Phys. Rev. Lett.}, 80:2245, 1998.

\bibitem{wurflinger2012nonlocal}
L.~E. W{\"u}rflinger, J.-D. Bancal, A.~Ac{\'\i}n, N.~Gisin, and T.~V{\'e}rtesi.
\newblock Nonlocal multipartite correlations from local marginal probabilities.
\newblock {\em Phys. Rev. A}, 86:032117, 2012.

\bibitem{Combing}
D.~Yang and J.~Eisert.
\newblock Entanglement combing.
\newblock {\em Phys. Rev. Lett.}, 103:220501, 2009.

\bibitem{yang2015bidirectional}
Z.~Yang, P.~Hayden, and X.-L. Qi.
\newblock Bidirectional holographic codes and sub-ads locality.
\newblock {\em J. High Energy Phys.}, 01:175, 2015.

\bibitem{zeng2015quantum}
B.~Zeng, X.~Chen, D.-L. Zhou, and X.-G. Wen.
\newblock Quantum information meets quantum matter--from quantum entanglement
  to topological phase in many-body systems.
\newblock 2015.
\newblock arXiv:1508.02595.

\bibitem{zhang1998characterization}
Z.~Zhang and R.~W. Yeung.
\newblock On characterization of entropy function via information inequalities.
\newblock {\em IEEE Trans. Inf. Th.}, 44:1440--1452, 1998.

\bibitem{zhu2016clifford}
H.~Zhu, R.~Kueng, M.~Grassl, and D.~Gross.
\newblock The {Clifford} group fails gracefully to be a unitary 4-design.
\newblock 2016.
\newblock arXiv:1609.08172.

\end{thebibliography}

{\small\printindex}

\end{document}